\documentclass[useAMS,usenatbib]{mn2e}
\usepackage{graphicx}
\usepackage{subfigure}
\usepackage{times}
\usepackage{txfonts}
\usepackage{array}
\usepackage{caption}
\usepackage{threeparttable}
\usepackage{makecell}
\usepackage{color}
\usepackage{flushend}
\usepackage{float}

\title[Modelling the NLRs in SDSS I. sample statistics]{Modelling narrow-line regions of active galaxies in the Sloan Digital Sky Survey - I. Sample selection and physical conditions}
\author[Z.-T. Zhang, Y.-C. Liang and F. Hammer]{Zhitai Zhang$^{1,2}$\thanks{E-mail:
ztzhang@nao.cas.cn}, Yanchun Liang$^{1}$ and Fran\c{c}ois Hammer$^{3}$\\
$^{1}$Key Laboratory of Optical Astronomy, National Astronomical Observatories, CAS, 20A Datun Road, 100012 Beijing, PR China\\
$^{2}$University of Chinese Academy of Sciences, 19A Yuquan Road, Shijingshan District, 100049, Beijing, China\\
$^{3}$GEPI , Observatoire de Paris, CNRS, University Paris Diderot, 5 place Jules Janssen, 92195 Meudon, France}
\begin{document}

\date{Accepted 2012 December 24. Received 2012 December 24; in original form 2012 April 17}

\pagerange{\pageref{firstpage}--\pageref{lastpage}} \pubyear{2013}

\maketitle

\label{firstpage}

\begin{abstract}
    Using spectroscopy from the Sloan Digital Sky Survey Data Release Seven, we systematically determine the electron density $n_e$ and electron temperature $T_e$ of active galaxies and star-forming galaxies while mainly focusing on the narrow-line regions (NLRs). Herein active galaxies refer to composites, low-ionization narrow emission-line regions (LINERs) and Seyferts following the Baldwin-Phillips-Terlevich diagram classifications afforded by the SDSS data. The plasma diagnostics of $n_e$ and $T_e$ are determined through the \textit{I}[S {\small II}] $\lambda$6716/$\lambda$6731 and \textit{I}[O {\small III}] $\lambda$5007/$\lambda$4363 ratios, respectively. By simultaneously determining $n_e$ from [S {\small II}] and $T_e$ from [O {\small III}] in our [O {\small III}] $\lambda$4363 emission sample of 15 019 galaxies, we find two clear sequences: $T_{LINER}$ $\gtrsim$ $T_{composite}$ $>$ $T_{Seyfert}$ $>$ $T_{star-forming}$ and $n_{LINER}$ $\gtrsim$ $n_{Seyfert}$ $>$ $n_{composite}$ $>$ $n_{star-forming}$. The typical range of $n_e$ in the NLRs of active galactic nuclei (AGNs) is $10^{2-3}cm^{-3}$. The temperatures in the NLRs range from 1.0 to 2.0$\times$10$^4$ K for Seyferts, and the ranges were even higher and wider for LINERs and composites. The transitions of $n_e$ and $T_e$ from the NLRs to the discs are revealed.

    We also present a comparative study, including stellar mass (M$_\star$), specific star formation rate (SFR/M$_\star$) and plasma diagnostic results. We propose that $Y_{L}$ $\gtrsim$ $Y_{SY}$ $>$ $Y_{C}$ $>$ $Y_{SF}$, where $Y$ is the characteristic present-day star-formation time-scale. One remarkable feature of the Seyferts shown on an M$_\star$-SFR/M$_\star$ diagram, which we call the evolutionary pattern of AGN with high ionization potential, is that the strong [O {\small III}] $\lambda$4363 Seyferts distribute uniformly with the weak Seyferts, definitely a reverse of the situation for star-forming galaxies. It is a natural and well-known consensus that strong [O {\small III}] $\lambda$4363 emissions in star-forming galaxies imply young stellar populations and thus low stellar masses. However, in the AGN case, several strong lines of evidence suggest that some supplementary energy source(s) should be responsible for high ionization potential.

\end{abstract}


\section{Introduction}

    The investigation of the narrow-line regions (NLRs) of active galactic nuclei (AGN) is important for a number of reasons. The NLR provides a bridge from the familiar and relatively well understood to the exotic and poorly known. On one hand, this region connects on its inner side with the more compact emitting regions that are spatially unresolved (except in the radio) and occur in very obscure surroundings. On the other hand, the outer boundary of the NLR does not just conjoin with the interstellar medium (ISM) of the galaxy but it can often be well spatially resolved at optical wavelengths. \citet{ferland} were the first to attempt to model the NLRs of AGNs, and many other studies followed. Emission-line region models are used to determine the physical conditions of the ionized gas (for details, refer to \citealt{dopita1}). Currently, two main ionization sources are most frequently applied: photoionization and shock ionization (refer to an overview of modelling NLRs in \citealt{groves1}). In addition to the ionization status, the geometry, spatial extent and emission from this region could also be explained and reproduced by modelling the NLRs. However, before running the models, it should be noted that all of these types of models rely on pre-estimates of some input parameters (e.g. density and temperature). Fortunately, the emission lines observed in the NLRs of AGNs are much the same as those observed in H {\small II} regions and planetary nebulae. Therefore, the standard nebular diagnostic methods can be applied to determine the electron density and temperature in the NLRs of AGNs (see \citealt{osterbrock1} for more details). One of the earliest and most well-studied examples is the NLR in Cyg A. The amount of extinction calculated from the Balmer decrement giving the best overall fit with the recombination decrement for \textit{$T_e$}$=$10$^4$ K, \textit{$n_e$} $=$ 10$^4$ cm$^{-3}$ in Cyg A (\citealt{osterbrock2}; \citealt{osterbrock3}). \citet{koski} have found that the average density in the line-emitting gas of 20 Seyfert 2 galaxies is approximately 2000 cm$^{-3}$ and that the average temperature is approximately 12 000 $\sim$ 25 000 K. \citet{walsh} has studied the Seyfert nuclei of NGC 1068 and has found that the electron density range is 9$\times$10$^4$ to 5$\times$10$^5$ cm$^{-3}$, assuming an electron temperature of $\sim$15 000 K.

    Two different types of strategies are often applied to investigate an object class: one is to study one or a few well-observed individual objects, and the other is to study a large sample of objects. The advantage of the former strategy is the more detailed and accurate treatment of data, yet it poses the risk of misidentifying the truly representative characteristics. In contrast, the latter strategy, which is usually based on a homogeneous data-set, allows us to utilize the information contained in collective trends and correlations. The former strategy has been more frequently used when investigating the physical conditions of NLRs (e.g. \citealt{kraemer1}; \citealt{kalser}; \citealt{barth}; \citealt{collins}; \citealt{kraemer2}). In comparison, there have been only a few previous estimates of the density and temperature of a large sample of AGN NLRs (e.g. \citealt{rodriguez}; \citealt{sulentic}; \citealt{bennert}; \citealt{xu}). Because of the success of the Sloan Digital Sky Survey (SDSS), we have a large data set to use in order to investigate the NLRs of AGNs in the local Universe (e.g. \citealt{hao1,hao2}; \citealt{zhou}; \citealt{liu}; \citealt{shen}). In order to prepare accurate and trustable input parameters for the next-step modelling, we follow the second strategy in this study.

    Narrow emission-line galaxies can be generally separated into four classes: star-forming galaxies, composites, low-ionization nuclear emission-line region (LINERs), and Seyfert galaxies. It is important to understand how these four classes differ in their ionization sources and thus temperatures, densities as well as emission, in order to build a picture of the local Universe. \citet{baldwin2} (hereafter BPT) have proposed using optical emission-line ratios to classify the dominant energy source in emission-line galaxies. The BPT diagrams are sensitive to the hardness of the ionizing radiation field. The first semi-empirical classification to be used with the standard optical diagnostic diagrams was derived by \citealt{osterbrock4} and \citealt{veilleux}, providing better optical classifications. These classification schemes have been significantly improved using the SDSS (e.g. \citealt{kewley3}, hereafter Ke06). With the improved optical classifications, for the first time, we are able to analyse the plasma diagnostic results accurately and to systematically distinguish between the different classes within an observationally homogeneous sample.

    The present paper is organised as follows. In Section 2, we give a description of our sample selection criteria. Sample definitions are included in Section 3, while Section 4 is devoted to the plasma diagnostic results. In Section 5, we focus on NLRs, providing statistical $n_e$ and $T_e$ analyses and making comparisons between different classes. In Section 6, we further apply the diagnostic results obtained and present a comparative study dealing with the relationships of physical properties, such as stellar mass (M$_\star$) and specific star formation rate (SFR/M$_\star$), in different classes. In Section 7, we discuss the shock effects and low-metallicity AGN candidates for further study. Finally, we summarize our main results in Section 8. Throughout this paper, we assume a flat $\Lambda$CDM cosmology with $\Omega_m$ = 0.3, $\Omega_{\Lambda}$ = 0.7, and $H_o$ = 70 km s$^{-1}$ Mpc$^{-1}$.

\section{Sample Selection}

    The SDSS (\citealt{york}; \citealt{stoughton}) uses a dedicated, wide-field, 2.5-m telescope \citep{abazajian} at the Apache Point observatory, New Mexico, for CCD imaging in five broad-bands, \textit{ugriz}, over 10 000 deg$^2$ of high-latitude sky, and for the spectroscopy of a million galaxies and 100 000 quasars over this same region; these goals have been realized with the seventh public data release (DR7\footnote{http://www.sdss.org/dr7}; \citealt{abazajian}).
    The images were reduced (\citealt{lupton2}; \citealt{stoughton}; \citealt{pier}; \citealt{ivezic}) and calibrated (\citealt{hogg}; \citealt{smith}; \citealt{tucker}), and galaxies were selected in two ways for follow-up spectroscopy covering 3800-9200 {\AA} at a resolution of $\lambda/\Delta\lambda\simeq$ 1800. The spectra were taken using 3-arcsec diameter fibres, positioned as close as possible to the centres of the target galaxies. Several improvements have been made to the spectroscopic software since DR5, particularly with regards to wavelength calibration, spectrophotometric calibration and the handling of strong emission lines (\citealt{adelman}; \citealt{abazajian}).

    The SDSS data base has been explored by several groups, using different approaches and techniques. Our analysis is based on data products from the catalogues\footnote{Available at http://mpa-garching.mpg.de/SDSS/DR7} obtained by the Max-Planck Institute for Astronomy (MPA, Garching) and John Hopkins University (JHU). The MPA/JHU catalogues are excellent resources for statistical studies of nearby narrow-line AGN populations because of the very large sample size (e.g. 33 589 AGN in DR2 and 88 178 in the DR4 catalogue) and because of the inclusion of additional data for the host galaxies, such as emission-line analyses and some derived physical properties, including stellar masses (\citealt{kauffmann1}; \citealt{gallazzi}; \citealt{salim}), star formation rates \citep{brinchmann}, and gas phase metallicities for star-forming galaxies \citep{tremonti}. The MPA/JHU group has publicly released catalogues for a total of 927 552 SDSS galaxies corresponding to DR7, a significant increase in size from their previous DR4 release. A number of improvements are made due to develppments both in the SDSS reduction and analysis pipelines. The emission-line fluxes given in the MPA/JHU catalogues have all been corrected for foreground (galactic) reddening \citep{o'donell}.

\subsection{Selection criteria}

    Our sample selection criteria are as follows.

    \begin{enumerate}
     \item The signal-to-noise (S/N) is $>$ 5 in the strong emission lines [O {\small II}] $\lambda$3726, 29 {\AA}, H$\beta$ $\lambda$4861 {\AA}, [O {\small III}] $\lambda$5007 {\AA}, [O {\small I}] $\lambda$6300 {\AA}, H$\alpha$ $\lambda$6563 {\AA}, [N {\small II}] $\lambda$6584 {\AA}, [S {\small II}] $\lambda$6716, 31 {\AA}. The S/N criteria are required so that we can accurately classify the galaxies into star-forming or AGN-dominated classes. As discussed on the website\footnote{http://www.mpa-garching.mpg.de/SDSS/DR7/raw\_data.html}, the uncertainties (i.e. $^*\_\verb"FLUX"\_\verb"ERR"$) are likely to be underestimates of the true uncertainties. By using the duplicate observations of galaxies to compare the empirical spread in value determinations with the random errors, the MPA/JHU group presents the results given as `scale uncertainties' in their DR7 release to correct for the underestimates. We adopt this correction and increase the uncertainty estimates on the emission lines by multiplying by the listed line flux uncertainty estimate factors.

     \item Objects with no stellar mass (M$_\star$) measurements (both total and fibre estimates) and poor specific star formation rate (SFR/M$_\star$) estimation (i.e. $\verb"FLAG"$ = 1) in the catalogues are excluded. We utilize these two derived parameters to probe the galaxy formation histories of different classes of emission-line galaxies in our sample.

     \item Broad-line AGN contaminations are removed. \citet{hao1} have discussed the method of identifying broad-line AGNs in the SDSS DR4 spectra. They have proposed FWHM(H$\alpha$) $>$ 1200 km s$^{-1}$ as the selection criterion for defining broad-line AGNs. We apply this classification to our data to remove broad-line AGN contaminations, using the $\verb"SIGMA_BALMER"$ given in the MPA/JHU catalogues to derive FWHM$_{Balmer} = \verb"SIGMA_BALMER" \times$ 2.355.
    \end{enumerate}

    Note that herein we have not included any cuts on redshift, as in previous works on SDSS emission-line galaxies (e.g. Ke06). In fact, the aperture effects, in Section 5, we present and discuss the aperture effects, or more precisely how the determined $n_e$ and $T_e$ (plasma diagnostics in Section4) vary as a function of the physical aperture size. Although LINERs typically have lower luminosities than Seyfert galaxies and are therefore found at lower redshifts than Seyferts in the magnitude-limited SDSS, this issue does not strongly affect our analysis because we are only interested in the typical range and representative values of $n_e$ and $T_e$.

    After applying the above three criteria, our resulting parent sample contains 46 867 emission-line galaxies.

\subsection{Extinction correction}

    The emission line fluxes are corrected for dust extinction. In this study, we use the assumptions given by previous studies. We use the Balmer decrement for Case B recombination at $T = 10^4$ K and $n_e \sim 10^{2-4}$ cm$^{-3}$, the dust-free Balmer-line ratio H$\alpha$/H$\beta$ value of 2.86 for star-forming galaxies and theoretical AGN value of 3.1 for composite galaxies, LINERs and Seyferts, and the relation given by \citet{osterbrock1}

    \begin{equation}   {\left(\frac{I_{H\alpha}}{I_{H\beta}}\right)_{obs}=\left(\frac{I_{H\alpha0}}{I_{H\beta0}}\right)_{intr}10^{-c(f(H\alpha)-f(H\beta))}}.
    \label{extinction}
    \end{equation}
    Applying the average interstellar extinction law, we have $f(H\alpha)-f(H\beta)=0.347$ as summarized from \citet{fitzpatrick}.

    Theoretically, the Balmer decrement has been determined over a range of temperatures and densities for two cases: Case A (assuming the Lyman series is optically thin) and Case B (assuming the Lyman series is optically thick). Generally, Case B recombination is a better approximation for most nebulae.  Some calculated Case B results (i.e. ratio descriptions) for the H I recombination lines have been given by table B.7 in \citet{dopita1} or table 4.4 in \citet{osterbrock1}.  Although we can conversely obtain observational Balmer decrements with Balmer series measurements using $n_e$ and $T_e$ estimates and the given ratio descriptions, the variations as a result of temperature and density should not be significant compared with our current values. Furthermore, although we can predict the H$\alpha$/H$\beta$ value for Case B by using the $T_e$ and $n_e$ estimates, considerable errors must be introduced by not only the measurements of line intensity uncertainties but also the algorithm used for solving $n_e$ and $T_e$. We should mention here that \citet{groves3} have found a systematic bias in the SDSS DR7 MPA/JPU catalogues (i.e. a 0.35-{\AA} underestimation of the emission-line H$_\beta$ equivalent width) and have suggested that the DR7 $A_V$ estimates could be overestimated by a mean difference of - 0.07 mag. However, this effect is not important in our analysis.

\section{Sample definition}

\subsection{Sample: [O III] $\lambda$4363 detection quality}

    Precise $T_e$ measurements require careful CCD calibrations, along with fairly high-quality spectra. In practice, however, the [O {\small III}] $\lambda$5007/$\lambda$4363 ratio is quite large and is thus difficult to measure accurately. Therefore, we further reject 31 150 objects with poor [O {\small III}] $\lambda$4363 detection quality (i.e. S/N $<$ 1). The remaining 15 717 objects are divided into the three samples shown in Table \ref{SN}: strong (sample A), intermediate (sample B), and weak (sample C) [O {\small III}] $\lambda$4363 emission samples, respectively. As expected, we see that stronger [O {\small III}] $\lambda$4363 objects display significantly higher luminosities of [O {\small III}] $\lambda$5007 ($L_{\tiny {\textrm{[O III]}}}$).

    \begin{table}
     \centering
     \begin{threeparttable}
     \caption{Sample divisions based on [O {\small III}] $\lambda$4363 detection quality. Fluxes summarized here are pre-extinction correction measurements of the [O {\small III}] $\lambda$4363 line (in units of 10$^{-17}$ erg s$^{-1}$ cm$^{-2}$). Luminosities listed are in units of $L_\odot$ after extinction correction. Note that $L_{\tiny {\textrm{[O III]}}}$ is the luminosity of [O {\small III}] $\lambda$5007.}
     \label{SN}
        \begin{tabular}{l c c c}
        \hline\\[-1.4 ex]
         & Sample A & Sample B & Sample C \\
         \\[-1.4 ex]
        \hline\\[-2 ex]
         Number & 1098 & 1409 & 13 210 \\
         S/N & $>$5 & 3-5 & 1-3 \\
         \emph{z}$_{median}$ & 0.070 & 0.067 & 0.075 \\
         \emph{z}$_{range}$ & 0.03-0.32 & 0.02-0.29 & 0.02-0.32 \\
         Flux$_{median}$ & 31.0 & 13.3 & 5.0 \\
         Flux$_{range}$ & 6.2-480.9 & 4.9-66.1 & 1.1-50.1 \\
         \\[-1.4 ex]
         log $L_{\tiny {\textrm{[O III]}}}$ & 8.32 & 8.01 & 7.47 \\
         log $L_{\tiny {\textrm{H$\beta$}}}$ & 7.59 & 7.45 & 7.47 \\
         \\[-2 ex]
        \hline
        \end{tabular}
        \end{threeparttable}
     \end{table}

    The three most widely used BPT diagrams are applied to classify the emission-line galaxies: [N {\small II}] $\lambda$6584/H$\alpha$ versus [O {\small III}] $\lambda$5007/H$\beta$ (hereafter [N {\small II}]/H$\alpha$ versus [O {\small III}]/H$\beta$), [S {\small II}] $\lambda$6584/H$\alpha$ versus [O {\small III}] $\lambda$5007/H$\beta$ (hereafter [S {\small II}]/H$\alpha$ versus [O {\small III}]/H$\beta$) and [O {\small I}] $\lambda$6584/H$\alpha$ versus [O {\small III}] $\lambda$5007/H$\beta$ (hereafter [O {\small I}]/H$\alpha$ versus [O {\small III}]/H$\beta$) (e.g. \citealt{veilleux}). Using the calibrations obtained by \citet{kewley1} (hereafter Ke01), \citet{kauffmann3} (hereafter Ka03) and Ke06, we subdivide each of the three samples into four classes: star-forming galaxies, composites, LINERs, and Seyferts.

    \begin{figure}
     \centering
     \includegraphics[scale=0.47]{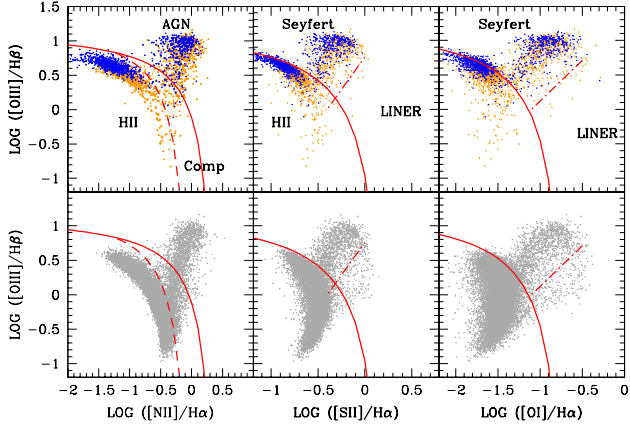}
     \\
     \caption{Three BPT diagrams (from left to right): the [N II]/H$\alpha$ versus [O III]/H$\beta$ diagram, the [S II]/H$\alpha$ versus [O III]/H$\beta$ diagram and the [O I]/H$\alpha$ versus [O III]/H$\beta$ diagram. Top panels: sample A (blue, upper) and sample B (orange, lower) are illustrated. Bottom panels: sample C objects (grey) are presented.  The Ke01 extreme starburst classification line (solid line), the Ka03 pure star formation line (dashed line) and the Ke06 Seyfert-LINER line (dot-dashed line) are used to separate galaxies into four classes: star-forming galaxies, composites, Seyferts and LINERs.}
    \label{FigBPT}
    \end{figure}

    According to this scheme, our sample of 15 717 galaxies contains 1842 (11.7 per cent) Seyferts, 95 (0.6 per cent) LINERs, 1338 (8.5 per cent) composites, and 11 744 (74.7 per cent) star-forming galaxies. The rest are ambiguous galaxies (698; 4.4 per cent), which are classified as one type of object in one or two diagrams and classified as another type of object in the remaining diagram(s). Hereafter we exclude the 698 ambiguous galaxies. Fig. \ref{FigBPT} shows the sample distributions in the three BPT diagrams. It is confirmed that (strong) [O {\small III}] $\lambda$4363 is more easily detected for Seyferts, which is also a clue for some alternative source of ionization in these galaxies. The compositions of the different galaxy classes in sample A, B and C are shown in Table. \ref{Luminosity}. It is clear that $L_{\tiny {\textrm{[O III]}}}$ shows a correlation with the [O {\small III}] $\lambda$4363 intensity, an effect that is not seen in $L_{\tiny {\textrm{H$\beta$}}}$. Seyferts, LINERs and composites have similar redshift values, and are more distant than star-forming galaxies (refer to Table. \ref{Luminosity}). In addition, Seyferts and LINERs have higher $L_{\tiny {\textrm{[O III]}}}$ than composites and star-forming galaxies, which suggests different ionization sources are present in different type of galaxies.

    \begin{table}
     \caption{Number distribution of the subsamples discussed in the text, and a summary of the mean redshift and luminosity (in units of $L_\odot$).}
     \label{Luminosity}
     \centering
        \begin{tabular}{l l c c c c}
        \hline
         \\[-1.4 ex]
         & \multicolumn{3}{c}{Object} & \multicolumn{2}{c}{Luminosity} \\
         & Subsample & Number & Redshift & log $L_{\tiny {\textrm{[O III]}}}$ & log $L_{\tiny {\textrm{H${\beta}$}}}$ \\
         \\[-1.4 ex]
        \hline
         \\[-2 ex]
         ASY & Seyfert & 371 & 0.098 & 8.64 & 7.75 \\
         AL & LINER & 1 & 0.053 & 7.66 & 7.25 \\
         AC & composite & 33 & 0.107 & 8.35 & 7.85 \\
         ASF & star-forming & 529 & 0.071 & 8.08 & 7.44 \\
         \\[-1.4 ex]
         BSY & Seyfert & 419 & 0.096 & 8.41 & 7.55 \\
         BL & LINER & 5 & 0.107 & 8.46 & 8.01 \\
         BC & composite & 82 & 0.099 & 8.07 & 7.78 \\
         BSF & star-forming & 795 & 0.073 & 7.81 & 7.34 \\
         \\[-1.4 ex]
         CSY & Seyfert & 1052 & 0.100 & 8.24 & 7.52 \\
         CL & LINER & 89 & 0.085 & 7.74 & 7.40 \\
         CC & composite & 1223 & 0.094 & 7.79 & 7.83 \\
         CSF & star-forming & 10 420 & 0.077 & 7.37 & 7.42 \\
         \\[-2 ex]
        \hline
        \end{tabular}
     \end{table}

    \begin{table*}
     \centering
     \begin{threeparttable}
     \caption{Summary of A$_V$ in seven M$_\star$ bins. The superscripts show the sample size of those samples that have a sample size no larger than 10.}
     \label{Av_MassBin}
     \renewcommand\arraystretch{1.1}
        \begin{tabular}{l l l l l l l l l l l l l}
        \hline
         \Gape[5pt] log M$_\star$ (M$_\odot$) & \multicolumn{3}{c}{Seyferts} & \multicolumn{3}{c}{LINERs} & \multicolumn{3}{c}{Composites} & \multicolumn{3}{c}{Star-forming galaxies} \\
         \Gape[5pt]  & A & B & C & A & B & C & A & B & C & A & B & C \\
        \hline
         8.1 &  &  & 0.29$^1$ &  &  &¡¡&  &  &¡¡& 0.25 & 0.28 & 0.39 \\
         8.7 &  &  &  &  &  &  & 0.05$^2$ &  & 0.53$^3$ & 0.29 & 0.32 & 0.33 \\
         9.3 & 0.24$^7$ & 0.42$^3$ & 0.35$^3$ &  &  &  & 0.31$^5$ & 0.24$^4$ & 0.37$^8$ & 0.32 & 0.40 & 0.45 \\
         9.8 & 0.56 & 0.72 & 0.47 &  &  & 1.09$^5$ & 0.57$^3$ & 0.53 & 0.60 & 0.46 & 0.55 & 0.68 \\
         10.3 & 0.63 & 0.62 & 0.74 &  &  & 1.17 & 0.57 & 0.67 & 0.82 & 0.68 & 0.90 & 1.01 \\
         10.7 & 0.63 & 0.84 & 0.84 &  & 1.19$^3$ & 0.96 & 0.81$^{10}$ & 0.89 & 1.11 &  & 1.08$^{10}$ & 1.36 \\
         11.2 & 0.72 & 0.86 & 0.96 & 0.35$^1$ & 0.95$^2$ & 1.10 &  & 1.30$^3$ & 1.35 &  & 0.46$^1$ & 1.59 \\
        \hline
        \end{tabular}
     \end{threeparttable}
    \end{table*}

    After performing extinction corrections, we calculate $A_V$ in seven M$_\star$ bins (see Table. \ref{Av_MassBin}), where $A_V = R_V\times E(B-V) = (3.1c)/1.47$ where $R_V$ = 3.1, on average, in the Milky Way and $E(B-V) = c/1.47$ for the standard Galactic extinction law (refer to equation \ref{extinction} for $c$; \citealt{osterbrock1}). The following two relations are revealed: comparisons between samples A, B and C show $A_V$$_{-A}$ $<$ $A_V$$_{-B}$ $<$ $A_V$$_{-C}$; comparisons between the four classes of emission-line galaxies show $A_V$$_{-Seyfert}$ $<$ $A_V$$_{-Comp}$ $\lesssim$ $A_V$$_{-StarForming}$ $\lesssim$ $A_V$$_{-LINER}$ at constant M$_\star$. The first relation could be explained by a selection bias because it is far more difficult to observe faint lines, such as [O {\small III}] $\lambda$4363, in a heavily obscured medium. Considering the second relation, the sampled LINERs as a group have the highest $A_V$. Because of supernova remnants, the high $A_V$ in LINERs most likely indicates the coexistence of old stellar populations and some specific high-energy process.

\subsection{Class: NLR-dominated objects}

     The NLR is generally a few kiloparsecs across (e.g. \citealt{bennert0}; \citealt{netzer}), although ionized gas has been seen at even larger ranges in some objects. Therefore, the NLR is a limited-small region in its host galaxy, and is most likely a combination of gas fluorescing in response to the active nuclei and material being ionized by massive stars nearby. The SDSS 3-arcsec diameter fibres subtend a variable amount of the total light as the redshift varies; therefore, the aperture effects are significant when considering physical conditions such as $n_e$ and $T_e$. In this section, we focus on the NLRs in our samples to gain insight into the relative contributions of the NLR and disc as the physical aperture sizes vary.

\begin{figure}
      \centering
      \includegraphics[scale=0.44]{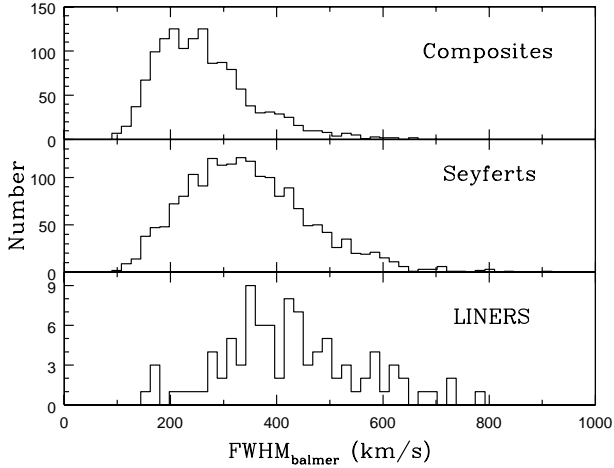}
      \\
      \caption{Histograms of FWHM of Balmer lines for the three types of emission-line galaxies (from top to bottom): composites, Seyferts and LINERs.}
      \label{FigFWHM}
     \end{figure}

     In the spectra of Seyfert 2 galaxies, the narrow (permitted and forbidden) emission lines show typical linewidths (FWHM) of 300-500 km s$^{-1}$ (e.g. \citealt{kollatschny}). For LINERs, the FWHM (permitted and forbidden) could be even greater than those of the Seyfert 2 galaxies. Fig. \ref{FigFWHM} shows histograms of FWHM$_{Balmer}$ for active galaxies (i.e. Seyferts, LINERs and composites; cf. the estimation of FWHM$_{Balmer}$ in Section 2). The linewidths range from 100 to 600-800 km s$^{-1}$  with median values of 250, 335, and 419 km s$^{-1}$ for composites, Seyferts, and LINERs, respectively. We also check the FWHM of forbidden lines and the median values (in the same order) are: 264, 341 and 439 km s$^{-1}$. Not surprisingly, the transition objects (i.e. composites) have narrower lines compared to LINERs, as is expected because of differences in their average Hubble types and the well-known dependence of nebular linewidth on bulge prominence (e.g. \citealt{ho1}). In addition, we see that LINERs have wider permitted lines than Seyferts. Table \ref{FWHM} summarizes the FWHM-selected objects in detail. It is not surprising that objects with FWHM$_{Balmer}$ $>$ 300 km s$^{-1}$ are much more easily found in LINERs and Seyferts than in composites. Here, we notice that there are 168 (with S/N $>$ 1; $\sim$1.4 per cent) narrow-line star-forming galaxies. Thus we plot these star-forming galaxies in the BPT diagram (i.e. [N {\small II}]/H$\alpha$ versus [O {\small III}]/H$\beta$) and we find that these objects generally lie along the Ka03 pure star-forming line with no significant dispersions; we suppose they could be misclassified and actually be transition objects (i.e. composites).

\begin{figure}
      \centering
      \includegraphics[scale=0.40]{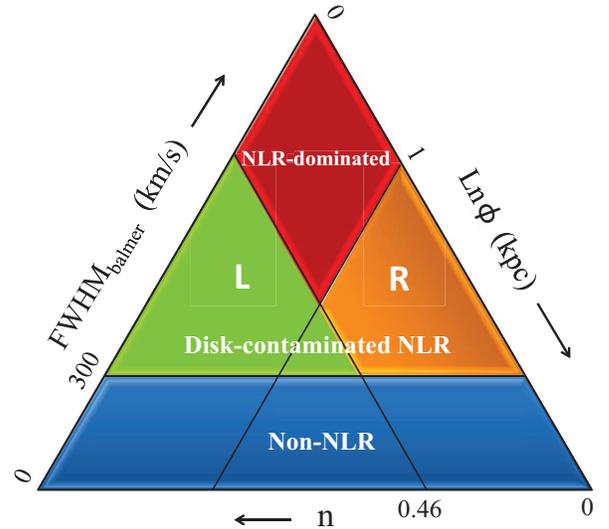}
      \\
      \caption{This ternary diagram illustrates selection criteria of the NLR-dominated objects (top red diamond). The disc-contaminated NLR objects occupy the left (L; green triangle) and right (R; orange trapezoidal) regions, while bottom blue trapezoid is occupied by the non-NLR objects. The arrows show scaling directions. Refer to equation (\ref{B02}) for the power law index $n$.}
      \label{FigNLR}
     \end{figure}

     \citet{bennert} have studied the NLRs in Seyfert 2 galaxies from spatially resolved optical spectroscopy. They have suggested that the lower limits of NLR size are 1-5 kpc and are limited by either the S/N data or the lack of a strong surrounding stellar ionization field. We convert redshifts into physical aperture size (i.e. $\phi$ in kpc) for the SDSS 3-arcsec fibre. \citet{ho2} has claimed that the NLRs in luminous Seyferts span from 50 pc to 1 kpc in radius. Additionally, \citet{bennert0} have measured the NLR sizes based on a sample of Seyferts. They have found a strong correlation between the NLR radius ($R_{NLR}$) and the [O {\small III}]$\lambda$5007 luminosity, which was scaled to the cosmology by \citet{netzer}, adopted in this paper. This can be written as

     \begin{equation}
     {R_{NLR} = 2.1L^{0.52\pm0.06}_{\tiny {\textmd{[O III]}},42} (\textrm{in kpc}}),
     \label{B02}
     \end{equation}
     where $L^{0.52\pm0.06}_{\tiny {\textmd{[O III]}},42}$ = $L_{\tiny {\textmd{[O III]}}}$ / 10$^{42}$ erg s$^{-1}$ (the uncertainty on the constant 2.1 kpc is of the order of 15 per cent). We adopt this correlation to constrain the upper limits of $R_{NLR}$ in order to compare with the sampled aperture size ($R_{obs}$). Fig. \ref{FigNLR} illustrates our NLR selection criteria. We require that ln$\phi$ $<$ 1 kpc, FWHM$_{Balmer}$ $>$ 300 km s$^{-1}$ and $R_{obs}$ $<$ $R_{NLR}$ for the so-called NLR-dominated (hereafter ND) objects. The other two classes are the disc-contaminated NLR (hereafter DN) and non-NLR (hereafter NN) objects. Table \ref{defineNLR} is a summary of the three classes of objects. We examine how $n_e$ and $T_e$ vary as velocity dispersion increases. The results agree with the suggestion given by \citet{heckman1} that the kinematics and physical conditions of the narrow-line emitting gas are not strongly coupled. We utilize the class definitions in Section 5.

     \begin{table}
     \caption{Summary of objects with FWHM$_{Balmer}$ $>$ 300 km s$^{-1}$.}
     \label{FWHM}
     \centering
        \begin{tabular}{l r r r r r}
        \hline
         \\[-1.4 ex]
         Object & \multicolumn{2}{c}{S/N $>$ 1} & \multicolumn{2}{c}{S/N $>$ 3} \\
         & Number & Per cent & Number & Per cent \\
         \\[-1.4 ex]
        \hline
         \\[-2 ex]
         Seyfert(1842) & 1141 & 61.9 & 540 & 29.3 \\
         LINER(95) & 82 & 86.3 & 6 & 0.63 \\
         Composite(1338) & 383 & 28.6 & 46 & 3.4 \\
         SF(11 744) & 168 & 1.4 & 28 & 0.2 \\
         \\[-2 ex]
        \hline
        \end{tabular}
     \end{table}

     \begin{table*}
     \centering
     \begin{threeparttable}
     \caption{Mean values of redshift, $A_V$, FWHM$_{Balmer}$ (FWHM$_B$; km s$^{-1}$) and luminosity ($L_\odot$) of the three classes: NLR-dominated (ND), disc-contaminated NLR (DNL and DNR; refer to Fig. \ref{FigNLR}) and non-NLR (NN) objects.}
     \label{defineNLR}
       \begin{tabular}{l c c c c c c c c c}
       \hline
        \Gape[7pt] Class & Number & Redshift & $A_V$ & FWHM$_B$ & log $L_{\tiny {\textrm{[O III]}}}$ & log $L_{\tiny {\textrm{H$\beta$}}}$ & Sample A & Sample B & Sample C \\
       \hline
       \\[-2 ex]
       ND$_{SY}$ & 75 & 0.033 & 1.03 & 386 & 8.25 & 7.41 & 30 & 25 & 20\\
       ND$_L$ & 2 & 0.033 & 2.16 & 497 & 8.69 & 8.09 &  &  & 2 \\
       ND$_C$ & 21 & 0.034 & 1.67 & 379 & 8.09 & 8.17 &  & 2 & 19 \\
       ND$_{SF}$ & 1 & 0.041 & 2.20 & 306 & 8.21 & 8.86 &  &  & 1 \\
       \\[-1.4 ex]
       DNL$_{SY}$ & 637 & 0.118 & 0.66 & 394 & 8.34 & 7.54 & 104 & 135 & 398 \\
       DNL$_L$ & 429 & 0.096 & 0.94 & 451 & 7.82 & 7.50 & 1 & 3 & 65 \\
       DNL$_C$ & 69 & 0.118 & 0.97 & 378 & 8.06 & 8.04 & 9 & 30 & 274 \\
       DNL$_{SF}$ & 11 & 0.140 & 1.20 & 351 & 8.30 & 8.47 & 3 & 13 & 118 \\
       \\[-1.4 ex]
       DNR$_{SY}$ & 313 & 0.108 & 1.11 & 449 & 9.00 & 8.16 & 139 & 107 & 183 \\
       DNR$_L$ & 49 & 0.076 & 1.66 & 494 & 8.66 & 8.46 &  & 2 & 9 \\
       DNR$_C$ & 134 & 0.102 & 1.75 & 373 & 9.00 & 8.95 & 2 & 3 & 44 \\
       DNR$_{SF}$ & 33 & 0.122 & 1.39 & 335 & 9.01 & 8.97 & 7 & 51 & 21 \\
       \\[-1.4 ex]
       NN$_{SY}$ & 701 & 0.084 & 0.66 & 234 & 7.95 & 7.24 & 98 & 152 & 451 \\
       NN$_L$ & 13 & 0.048 & 0.84 & 234 & 7.01 & 6.71 &  &  & 13 \\
       NN$_C$ & 955 & 0.087 & 0.90 & 218 & 7.68 & 7.70 & 22 & 47 & 886 \\
       NN$_{SF}$ & 11 576 & 0.075 & 0.68 & 151 & 7.45 & 7.41 & 519 & 777 & 10 280 \\
       \\[-2 ex]
       \hline
       \end{tabular}
       \end{threeparttable}
     \end{table*}

\section{Plasma diagnostic results}

    The two best examples frequently used to measure the electron density are [S {\small II}] $\lambda$6716, 31 and [O {\small II}] $\lambda$3726, 29. However, the typical linewidths (FWHM) in NLRs, ranging from 300 to 500 km s$^{-1}$ (e.g. \citealt{kollatschny}), are comparable to, or are even larger than, the separation of the two [O {\small II}] lines, approximately 300 km s$^{-1}$. Therefore, the [O {\small II}] intensity ratio, which is a good electron density diagnostic in H {\small II} regions and planetary nebulae, cannot be applied to AGNs \citep{osterbrock1}. The [S {\small II}] ratio has two practical advantages for the determination of $n_e$: for z $<$ 0.3 the two [S {\small II}] lines fall in the rest-frame optical range and are hence easily observed, and the lines have a small enough separation and are thus not sensitive to reddening correction. The [O {\small III}] $\lambda$5007/$\lambda$4363 ratio is sensitive to density as well as temperature. The 5007 and 4363 lines have different critical densities, and at densities of $\gtrsim$10$^6$ cm$^{-3}$, the 5007 line is heavily suppressed because of the collisional de-excitation of the $^1$D$_2$ level \citep{osterbrock2}. Because of the different critical densities of the line transitions, the [O {\small III}] ratio is no longer a valid thermometer in the high-density (i.e. $\gtrsim$10$^6$ cm$^{-3}$) regions usually seen in broad-line regions (BLRs; e.g. \citealt{ho2}). However, considering that the average physical aperture coverage of the SDSS 3-arcsec diameter fibres for galaxies at z $>$ 0.02 (cf. Table \ref{SN}) is much larger than the typical size scale of the BLRs, the [O {\small III}] ratio is therefore still a reasonable temperature estimator in our case. Moreover, although the [O {\small III}] ratio is relatively sensitive to reddening correction because of the significant line separation, the mean impact for temperature measurements is no more than 10 per cent in our sample. In this study we choose to derive the electron density and electron temperature using \textit{I}[S {\small II}] $\lambda$6716/$\lambda$6731 (hereafter \textit{R}[S {\small II}]) and \textit{I}[O {\small III}] $\lambda$5007/$\lambda$4363 (hereafter \textit{R}[O {\small III}]), respectively.

    The calculations of electron density and temperature are processed with the current atomic data\footnote{References are listed at http://www.noao.edu/noao/staff/shaw/nebular} using the \textsc{TEMDEN} task within the \textsc{IRAF}\footnote{\textsc{IRAF} is distributed by the National Optical Astronomy Observatories, which are operated by the Association of Universities for Research in Astronomy under cooperative agreement with the National Science Foundation.} \textsc{STSDAS} package\footnote{Available at http://www.stsci.edu/institute/software\_hardware/stsdas} \citep{shaw}. The task is based on the five-level atom program first developed by \citet{de robertis}, including diagnostics from a greater set of ions and emission lines, particularly those in the ultraviolet satellite (e.g. \emph{International Ultraviolet Explorer} and \emph{Hubble Space Telescope}) archives.

    We treat samples A, B and C in a uniform way, except for sample C, which had poor [O {\small III}] $\lambda$4363 detections. That is, first we apply the uncertainty measurements of [O {\small III}] $\lambda$4363 to calculate the upper limits of the \textit{R}[O {\small III}] ratio at a 1$\sigma$ uncertainty, before performing the plasma diagnostics. In other words, the sample C temperatures are correspondingly lower limits because higher \textit{R}[O {\small III}] ratios lead to lower $T_e$. Higher assumed temperatures correspond to lower electron densities, and vice versa. Our plasma diagnostics are calculated as follows.

    \begin{enumerate}
      \item The initial estimate of $n_e$  is obtained with the [S {\small II}] ratio assuming, for example, $T$=10 000 K (for those with 1.43 $<$ \textit{R}[S {\small II}] $<$ 1.47, some lower values of $T$ are assumed).
      \item This density is then used to derive $T_e$[O {\small III}]. If this differs from the initial value of $T$ by more than 10 per cent, a new value of $n_e$ is then derived using the $T_e$[O {\small III}] obtained.
    \end{enumerate}
    Note that because the density-sensitive tracer \textit{R}[S {\small II}] used here is easily excited in a range of physical environments, we suggest that the $n_e$ estimates should be considered as the mean values of the sampled regions because of the multi-phase nature of the ISM. Correspondingly, the value of $T_e$ derived from \textit{R}[O {\small III}] with $n_e$ from \textit{R}[S {\small II}] will be biased to higher values. Given $n_e$ $\leq$ 10$^5$ cm$^{-3}$, the offsets would generally be no more than 2000 K as long as $T_e$ $<$ 20 000 K (e.g. \citealt{osterbrock1}). Therefore, we claim that the diagnostics applied compose the most acceptable and feasible methodology for analysing a large SDSS sample such as ours.

    \begin{table}
     \centering
     \begin{threeparttable}
     \caption{Summary of objects that have '$T_e$ with [O {\small III}] and $N_e$ with [S {\small II}]' (TONS). The four cases are: TONS$_1$, the calculation using the two line ratios leads to no consistent result; TONS$_2$, \textit{R}[S II]$>$1.47, and thus there are no density results; TONS$_3$, \textit{R}[O {\small III}]$<$11.1, and thus there are no temperature results; TONS$_4$, both density and temperature determinations fail.}
     \label{TONS}
        \begin{tabular}{l c c c c c}
        \hline
        \\[-1.4 ex]
        & TONS$_1$ & TONS$_2$ & TONS$_3$ & TONS$_4$ & Total \\
        \\[-1.4 ex]
        \hline
        \\[-2 ex]
         ASY(371) & 12 & 1 & 1 &  & 14 \\
         AL(1) &  &  &  &  & \\
         AC(33) & 3 &  & 2 &  & 5 \\
         ASF(529) & 96 & 47 &  &  & 143\\
         Total(934) & 111 & 48 & 3 &  & 162 \\
         \\[-1.4 ex]
         BSY(419) & 17 & 5 & 1 &  & 23\\
         BL(5) &  & 1 &  &  & 1 \\
         BC(82) & 14 & 3 & 5 &  & 22 \\
         BSF(795) & 204 & 104 & 14 & 2 & 324 \\
         Total(1301) & 235 & 113 & 20 & 370\\
         \\[-1.4 ex]
         CSY(1052) & 51 & 37 &  &  & 88 \\
         CL(89) & 13 & 2 & 2 &  & 17 \\
         CC(1223) & 181 & 71 & 17 & 1 & 270 \\
         CSF(10 420) & 3462 & 2328 & 101 & 35 & 5926\\
         Total(12 784) & 3657 & 2438 & 120 & 36 & 6251\\
         \\[-2 ex]
         \hline
        \end{tabular}
        \end{threeparttable}
    \end{table}

    \begin{table*}
    \centering
    \begin{threeparttable}
    \caption{Summary of the physical conditions (density and temperature). Note that by applying our determination procedures, the sample C temperatures are the lower limits. \emph{N$_{s}$} denotes the number of objects with \textit{R}[S II] $>$ 1.47 (i.e. the upper limit for \textsc{TEMDEN} calculation using \textit{R}[S II]). \emph{N$_o$} denotes the number of objects with \textit{R}[O {\small III}] $<$ 11.1 (i.e. the lower limit of \textit{R}[O {\small III}] for \textsc{TEMDEN} calculation). \emph{N$_{1.5}$}(per cent) is the number of objects with 1.5 $<$ $T_e$ $<$ 2.0 (10$^4$ K) and the fraction of the corresponding subsamples. \emph{N$_{2.0}$}(per cent) is the number of objects with $T_e$ $>$ 2.0 (10$^4$ K) and the fraction of the corresponding subsamples.}
    \label{PhysicalConditions}
       \begin{tabular}{l c c c c c c c c c c c}
       \hline
       \\[-1.4 ex]
        & & \multicolumn{2}{c}{\textit{R}[S II]} & \multicolumn{2}{c}{Electron density} & \multicolumn{2}{c}{\textit{R}[O III]} & \multicolumn{4}{c}{Electron temperature} \\
        & & & &\multicolumn{2}{c}{$n_e$ (cm$^{-3}$)} & & & \multicolumn{4}{c}{$T_e$ 10$^4$ (K)} \\
        Object & Num. & Range & N$_{s}$ & Range & Median/Mean & Range & N$_{o}$ & Range & Median/Mean & N$_{1.5}$(per cent) & N$_{2.0}$(per cent) \\
        (1) & (2) & (3) & (4) & (5) & (6) & (7) & (8) & (9) & (10) & (11) & (12) \\
        \\[-1.4 ex]
       \hline
       \\[-2 ex]
       ASY & 371 & 0.81-1.47 & 1 & 6-1620 & 422/447 & 8.8-2E2 & 1 & 0.5-5.7 & 1.5/1.7 & 154(42) & 45(12) \\
       AL & 1 & 0.73 &  & 3959 &  & 14.2 &  & 5.4 &  &  & 1(100) \\
       AC & 33 & 1.06-1.45 &  & 2-624 & 168/204 & 7.3-2E2 & 2 & 1.0-7.5 & 1.9/2.2 & 7(21) & 13(39) \\
       ASF & 529 & 0.96-1.71 & 45 & 1-780 & 62/79 & 14.3-3E2 & 0 & 0.7-5.5 & 1.2/1.2 & 25(5) & 3($<$1) \\
       \\[-1.4 ex]
       BSY & 419 & 0.81-1.71 & 5 & 3-1465 & 332/360 & 8.3-2E2 & 1 & 0.7-8.1 & 1.5/1.6 & 173(41) & 38(9) \\
       BL & 5 & 0.83-1.48 & 1 & 76-2244 & 508/777 & 17.0-6E1 &  & 1.7-4.2 & 1.8/2.3 &  & 5(100) \\
       BC & 82 & 0.85-1.55 & 3 & 3-1301 & 181/230 & 2.9-3E2 & 5 & 0.9-7.5 & 2.0/2.2 & 16(20) & 39(48) \\
       BSF & 795 & 0.93-2.85 & 104 & 1-1108 & 42/69 & 1.2-3E2 & 16 & 0.6-8.2 & 1.1/1.3 & 36(5) & 31(4) \\
       \\[-1.4 ex]
       CSY & 1052 & 0.80-2.05 & 37 & 1-1499 & 216/256 & 20.7-7E3 &  & 0.4-2.6 & 1.1/1.1 & 69(7) & 6($<$1) \\
       CL & 89 & 1.00-1.49 & 2 & 3-609 & 172/187 & 6.7-7E2 & 2 & 0.6-6.5 & 1.3/1.7 & 14(16) & 22(25) \\
       CC & 1223 & 0.87-1.66 & 72 & 1-1410 & 134/169 & 6.0-9E3 & 18 & 0.4-8.9 & 1.3/1.6 & 244(20) & 244(20) \\
       CSF & 10 420 & 0.86-2.53 & 2361 & 1-1141 & 32/55 & 4.1-6E5 & 136 & 0.3-10.0 & 1.0/1.4 & 1104(10) & 1437(14) \\
       \\[-2 ex]
       \hline
       \end{tabular}
    \end{threeparttable}
    \end{table*}

    \begin{figure*}
    \centering
    \subfigure[]{\includegraphics[scale=0.8]{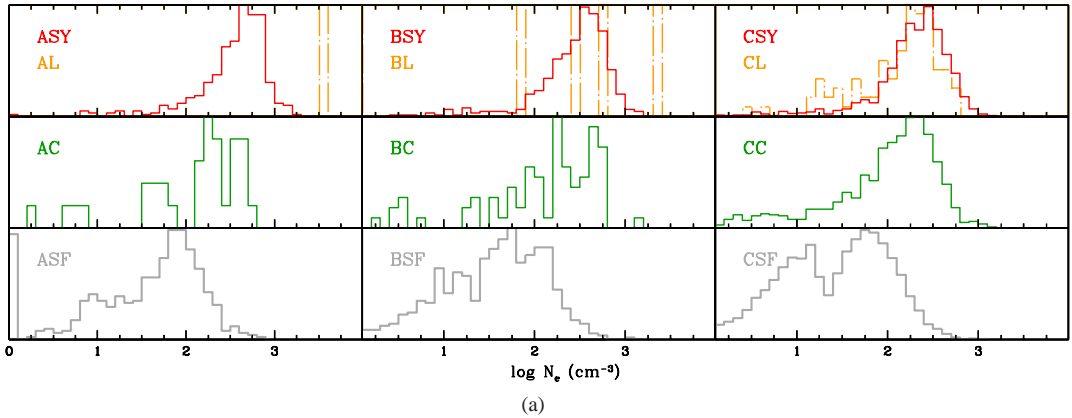}}
    \\
    \subfigure[]{\includegraphics[scale=0.8]{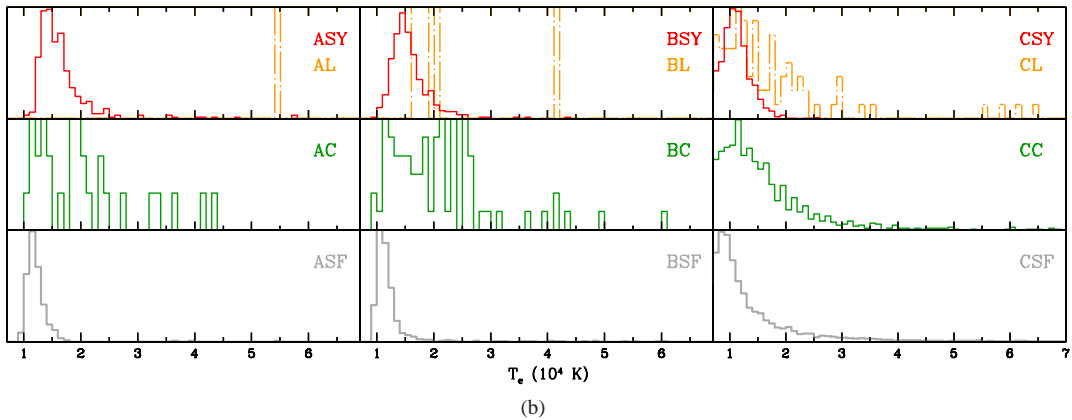}}
    \\
    \caption{Normalized histograms of electron density (a) and electron temperature (b). For each panel, from top to bottom, we show Seyferts (red solid) and LINERs (orange dot-dashed), composites (darkgreen), and star-forming galaxies (grey), and from left to right, we show samples A, B and C.}
    \label{FigTeNeHist}
    \end{figure*}

    At an assumed temperature or density, there are some objects labelled as 'TONS', which is short for '$T_e$ with [O {\small III}] and $N_e$ with [S {\small II}]'; this designation means that no (consistent) results were found using the simultaneous calculation processes (see Table \ref{TONS}). For example, it is noticeable that 2541 CSF (i.e. sample C star-forming galaxies) show $T_e$ $>$ 1.5$\times$10$^4$ K (cf. columns 11 and 12 of Table \ref{PhysicalConditions}), while 1742 ($\sim$ 70 per cent) of these are actually TONS$_1$ (1109) or TONS$_2$ (633) objects. Therefore, we further check the intensity ratios of these TONS objects.

    For the TONS$_2$, TONS$_3$ and TONS$_4$ objects, using the uncertainty measurements of the [S {\small II}] and [O {\small III}] lines to calculate the lower limits of \textit{R}[S {\small II}] and the upper limits of \textit{R}[O {\small III}], we find that $\sim$ 80 per cent of both TONS$_2$ and TONS$_3$ along with all TONS$_4$ objects show computable ratios. For the remaining 20 per cent of TONS$_2$ and TONS$_3$ objects, we suggest that the uncertainties introduced by [O {\small III}] extinction corrections have the greatest impact. Therefore, these three cases are most likely introduced by observational uncertainties. In the TONS$_1$ cases, we suggest an additional possibility, that the intrinsically different spatial distributions of the [S {\small II}] and [O {\small III}] emitting regions might lead to inconsistent results in the calculations of $n_e$ and $T_e$.

    The plasma diagnostic results are summarized in Table \ref{PhysicalConditions}, and Fig. \ref{FigTeNeHist} shows the normalized histograms of $n_e$ and $T_e$. The main results are as follows.

    \begin{description}
      \item[\emph{Electron density}.] Considering the density results, we can see a clear sequence (cf. columns 5 and 6 of Table \ref{PhysicalConditions}): $n_{LINER}$ $\gtrsim$ $n_{Seyfert}$ $>$ $n_{composite}$ $>$ $n_{star-forming}$. The typical $n_e$ uncertainties are 150 cm$^{-3}$ for Seyferts, LINERs and composites, and 80 cm$^{-3}$ for star-forming galaxies. Comparing samples A, B and C, the CSY and CL possess significantly lower $n_e$ ($\sim$150 cm$^{-3}$) than A/BSY and A/BL, most likely indicating some effect of shocks in strong [O {\small III}] $\lambda$4363 emitters (please refer to Table \ref{Luminosity} for the abbreviations). The characteristic $n_e$ range of active galaxies, including Seyferts, LINERs and composites, is 10$^{2-3}$ cm$^{-3}$, while that of the star-forming galaxies is 10$^{1-2}$ cm$^{-3}$. As shown in Fig. \ref{FigTeNeHist}(a), the $n_e$ sequence mentioned above is visible. However, we see a bimodal number distribution of $n_e$, which is especially significant for star-forming galaxies. To check whether this is a real effect, we examine the \textit{R}[S {\small II}] distribution and we can confirm no corresponding bimodality. In fact, the typical uncertainty for objects with $n_e$ $<$ 10 is $\sim$40 cm$^{-3}$ in our sample. As a result, the bimodal distribution is most likely introduced by the algorithm used for calculating the electron density in the \textsc{TEMDEN} package and is therefore not a physical real effect.
      \item[\emph{Electron temperature}.] The temperature results are listed in Table \ref{PhysicalConditions} (cf. columns 9 to 12). Fig. \ref{FigTeNeHist}(b) shows normalized temperature histograms for each galaxy type. As given above, sample C temperatures should be viewed as the lower limits for comparison. First, a $T_e$ sequence is clearly shown: $T_{LINER}$ $\gtrsim$ $T_{composite}$ $>$ $T_{Seyfert}$ $>$ $T_{star-forming}$. The typical uncertainties of $T_e$ for the four galaxy classes are (in units of 10$^4$ K) 0.2 for Seyferts, 0.3 for LINERs, 0.5 for composites and 0.1 for star-forming galaxies. Secondly, the sampled LINERs display a mean $T_e$ of approximately 2$\times$10$^4$ K (cf. column 10 of Table \ref{PhysicalConditions}) and show the highest number fraction of high $T_e$ species (i.e. $T_e$ $>$ 1.5$\times$10$^4$ K; cf. column 11 and 12 of Table \ref{PhysicalConditions}). Thirdly, a remarkable result is that the composites as a group show a value of $T_e$ very close to that of the LINERs and approximately 5000 K higher than that of the Seyferts. This result reveals the existence of very different dominating ionizing source(s) in composites from those in star-forming galaxies. For star-forming galaxies, the derived temperatures are generally consistent with the classical diagnostic results of bright planetary nebula ($\sim$ 1.1$\times$10$^4$ K). Moreover, as revealed in Fig. \ref{FigTeNeHist}, the active galaxies (e.g. LINERs and composites) show a significant tail extending to much higher temperatures, up to 6$\times$10$^4$ K, which is far too high to be explained only by photoionization caused by stars.
    \end{description}

\section{Electron density and electron temperature in NLRs}

    \begin{figure}
      \centering
      \includegraphics[scale=0.49]{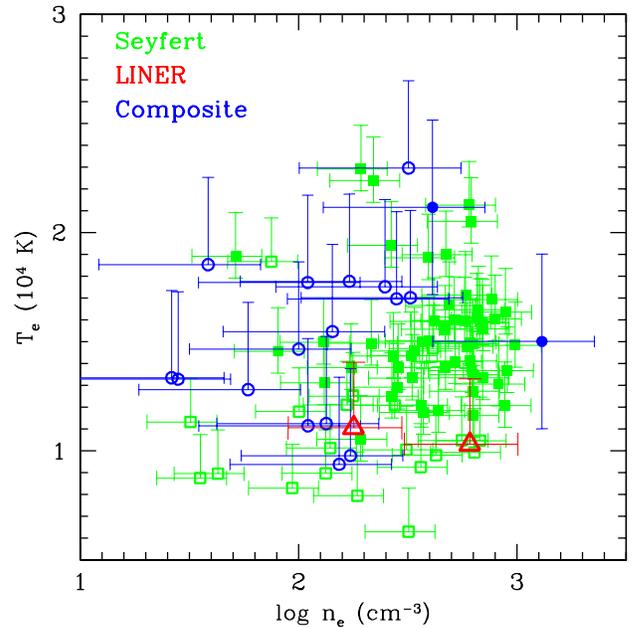}
      \\
      \caption{Electron density versus electron temperature. Sample distribution of the NLR-dominated objects: Seyferts (blue boxes), LINERs (red triangles)and composites (green circles), with error bars showing 1$\sigma$ dispersions. Open symbols denote objects from sample C with lower $T_e$ limits.}
      \label{FigTeNeDis}
    \end{figure}

    \begin{figure}
      \centering
      \includegraphics[scale=1.6]{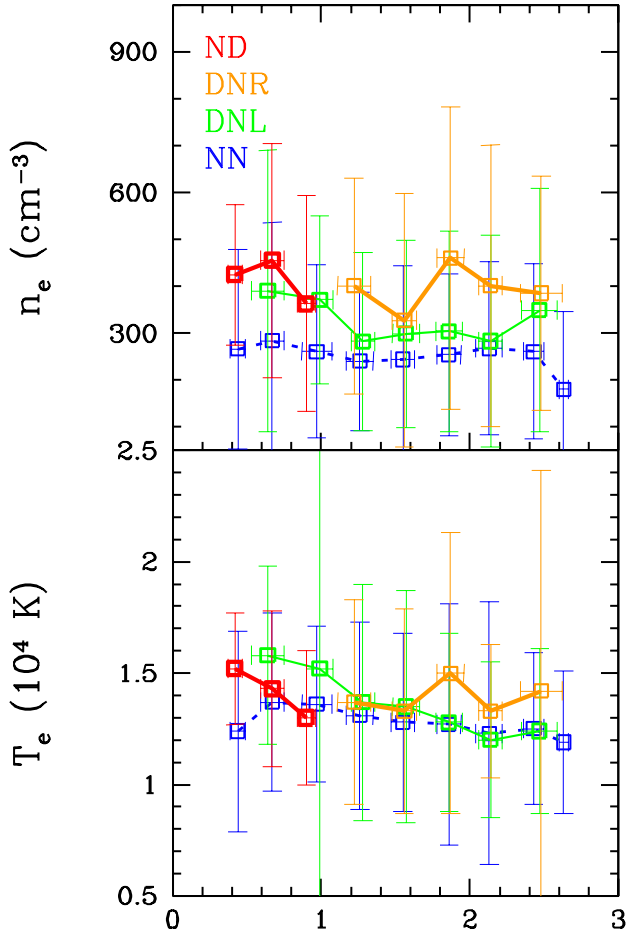}
      \\
      \caption{Physical aperture size ln$\phi$ versus $n_e$ (top) and $T_e$ (bottom) for Seyferts in bins of ln$\phi$. Empty boxes are the median values of ln$\phi$, $n_e$ and $T_e$, with error bars showing 1$\sigma$ dispersions: NLR-dominated (ND, red), disc-contaminated NLR (DNR, orange; DNL, green thin; refer to Fig. \ref{FigNLR}) and non-NLR objects (NN, dashed blue).}
      \label{FigPhyNT}
    \end{figure}

    \begin{description}
      \item[\emph{Electron density in NLRs}.] As shown in Fig. \ref{FigTeNeDis}, most NLRs lie in the $n_e$ range of 10$^{2-3}$ cm$^{-3}$. Only seven Seyferts, seven composites and one LINER have lower values of $n_e$, some of which might be the result of flux measurement uncertainties. We have tried higher densities of 10$^{5-6}$ cm$^{-3}$ to calculate $T_e$[O {\small III}] for our selected NLRs, but a significant fraction of Seyferts show values no greater than 6000 K, which could not be representative of the cases of NLRs of AGNs found in the literature (e.g. \citealt{bennert}). Thus, our results suggest that the characteristic $n_e$ range in NLRs of active galaxies is 10$^{2-3}$ cm$^{-3}$.
      \item[\emph{Electron temperature in NLRs}.] As illustrated in Fig. \ref{FigTeNeDis}, the typical range of $T_e$ in Seyferts is 1.0-2.0$\times$10$^4$ K, with a median value of 1.4$\times$10$^4$ K. Because both the selected LINERs and composites contain larger fractions of sample C objects than the Seyferts, and the $T_e$ derived could thus be biased to lower values, the typical range of these two classes could be higher and wider than that shown in Fig. \ref{FigTeNeDis}.
    \end{description}

    Fig. \ref{FigPhyNT} plots $n_e$ (top) and $T_e$ (bottom) as functions of physical aperture size ln$\phi$ for Seyferts in bins of ln$\phi$. For a precise evaluation, we further calculate the mean values of $n_e$ and $T_e$ for different types of objects. As shown in Table \ref{NLRvsHost}, the $n_e$ and $T_e$ of Seyferts show transitions from the NLR to the disc: $n_{ND_{SY}}$ $>$ $n_{DNR_{SY}}$ $>$ $n_{DNL_{SY}}$ $>$ $n_{NN_{SY}}$ and $T_{ND_{SY}}$ $>$ $T_{DNR_{SY}}$ $>$ $T_{DNL_{SY}}$ $>$ $T_{NN_{SY}}$ (refer to Table \ref{defineNLR} for the classifications). The unexpected low value of $T_e$ for NLR-dominated NLR LINERs and the high value of $T_e$ for non-NLR LINERs can most likely be explained by selection biases: the former is a result of the small sample size (i.e. only two from sample C) and the latter is because the non-NLR LINERs are sampled at lower redshifts than the disc-contaminated NLR LINERs and are biased to regions close to the galaxy centres. For composites and star-forming galaxies, we find no clear transition, as seen in the Seyferts. Therefore, we have checked further and found that this result was due to misclassification; that is, the correlation (see equation \ref{B02}) found by \citet{bennert0} was derived from Seyferts.

    We have reported single and statistical estimates of $n_e$ and $T_e$ for the NLRs of active galaxies. Physically, the emission-line regions are, by their nature, multi-phase media, and thus densities and temperatures are strongly coupled with non-uniformity. When modelling the NLRs, both density and temperature could vary with different physical conditions, and such non-uniformity or multi-phase nature should be considered (e.g. \citealt{binette}; \citealt{komossa}; \citealt{baskin}). We will leave this discussion to a companion paper, which will focus on modelling.

    \begin{table}
    \centering
    \begin{threeparttable}
     \caption{Mean $n_e$ and $T_e$ of the NLR-dominated (ND), the disc-contaminated NLR (DNL and DNR) and the non-NLR (NN) objects. Strong (sample A), intermediate (sample B) and weak (sample C) [O {\small III}]$\lambda$4363 objects are all included. For classifications, please refer to Fig. \ref{FigNLR}.}
     \label{NLRvsHost}
        \begin{tabular}{l r r r r r r r r}
        \hline
         \\[-1.4 ex]
         & \multicolumn{4}{c}{$n_e$ (cm$^{-3}$)} & \multicolumn{4}{c}{$T_e$ (10$^4$ K)} \\
         & ND & DNR & DNL & NN & ND & DNR & DNL & NN \\
         \\[-1.4 ex]
        \hline
         \\[-2 ex]
         SY & 427 & 394 & 306 & 255 & 1.40 & 1.39 & 1.28 & 1.29 \\
         L & 394 & 413 & 244 & 92 & 1.07 & 1.75 & 1.73 & 1.87 \\
         C & 197 & 166 & 151 & 167 & 1.54 & 1.63 & 1.71 & 1.61 \\
         \\[-1.4 ex]
         SF & 109 & 148 & 94 & 44 & 1.78 & 1.44 & 1.93 & 1.32 \\
         \\[-2 ex]
        \hline
        \end{tabular}
    \end{threeparttable}
    \end{table}

\subsection{High-$T_e$ NLRs in line-ratio diagnostic diagrams}

    One item remaining in the problem of modelling of NLR is the so-called temperature problem: the difficulty in reproducing high $T_e$[O {\small III}] with photoionization models. For instance, \citep{heckman} have claimed that $T_e >$ 20 000 K requires a source of energy in addition to photoionization. Traditionally, two different avenues have been explored when modelling the NLRs: photoionization and shock ionization. It is generally believed that photoionization is the dominant excitation mechanism in most AGNs, while the intrinsic defect of all shock models (e.g. \citealt{allen}) is that they require shocks throughout the NLR and that they cannot, alone, explain all NLR emission because shock signatures cannot always be observed. There have been several previous to solve the temperature problem from different perspectives (e.g. \citet{binette}). Here, we compare some of the current models with our selected high-$T_e$ object data.

    Line-ratio diagnostic diagrams are commonly used to visualize and compare models with observations. The ratios of emission lines plotted against one other, depending upon the emission lines chosen, can show clear relationships in density, metallicity, and the ionization mechanism. For example, [O {\small III}] $\lambda$5007/[O {\small II}] $\lambda$3726,29 (hereafter [O {\small III}]/[O {\small II}]) is theoretically independent of the chemical abundance and is a very good indicator of the ionization parameter. [O {\small I}] $\lambda$6300/H$\alpha$ (hereafter [O {\small I}]/H$\alpha$) is sensitive to the ionization parameter. [O {\small III}] $\lambda$5007/H$\beta$ (hereafter [O {\small III}]/H$\beta$) depends on all the parameters but is also frequently used as an indicator of the ionization parameter. [N {\small II}] $\lambda$6584/H$\alpha$ (hereafter [N {\small II}]/H$\alpha$) is mostly sensitive to the gaseous abundance. [S {\small II}] $\lambda$6716,31/H$\alpha$ (hereafter [S {\small II}]/H$\alpha$) is sensitive to the density. The ionization parameter $U$ is defined as

    \begin{equation}\label{U}
        U=\frac{Q_{tot}}{4{\pi}r^2nc}
    \end{equation}
    where $Q_{tot}$ is the total number rate of hydrogen ionizing photons isotropically emitted by the nucleus, $r$ is the representative distance of the emission-line clouds from the central source and $n$ = $n_H$ is the total hydrogen number density. We calculate the extinction correction effects on multiple line ratios: [N {\small II}]/H$\alpha$, [O {\small III}]/H$\beta$, [S {\small II}]/H$\alpha$, [O {\small I}]/H$\alpha$, [O {\small III}]/[O {\small II}] and the temperature-sensitive [O {\small III}] ratio. The [O {\small III}]/[O {\small II}] ratio is the most sensitive to extinction correction, showing a $\sim$30 per cent decrease after the extinction correction, and the temperature-sensitive [O {\small III}] ratio decreases less than 10 per cent.

\begin{figure*}
    \centering
    \subfigure[]{\includegraphics[scale=0.46]{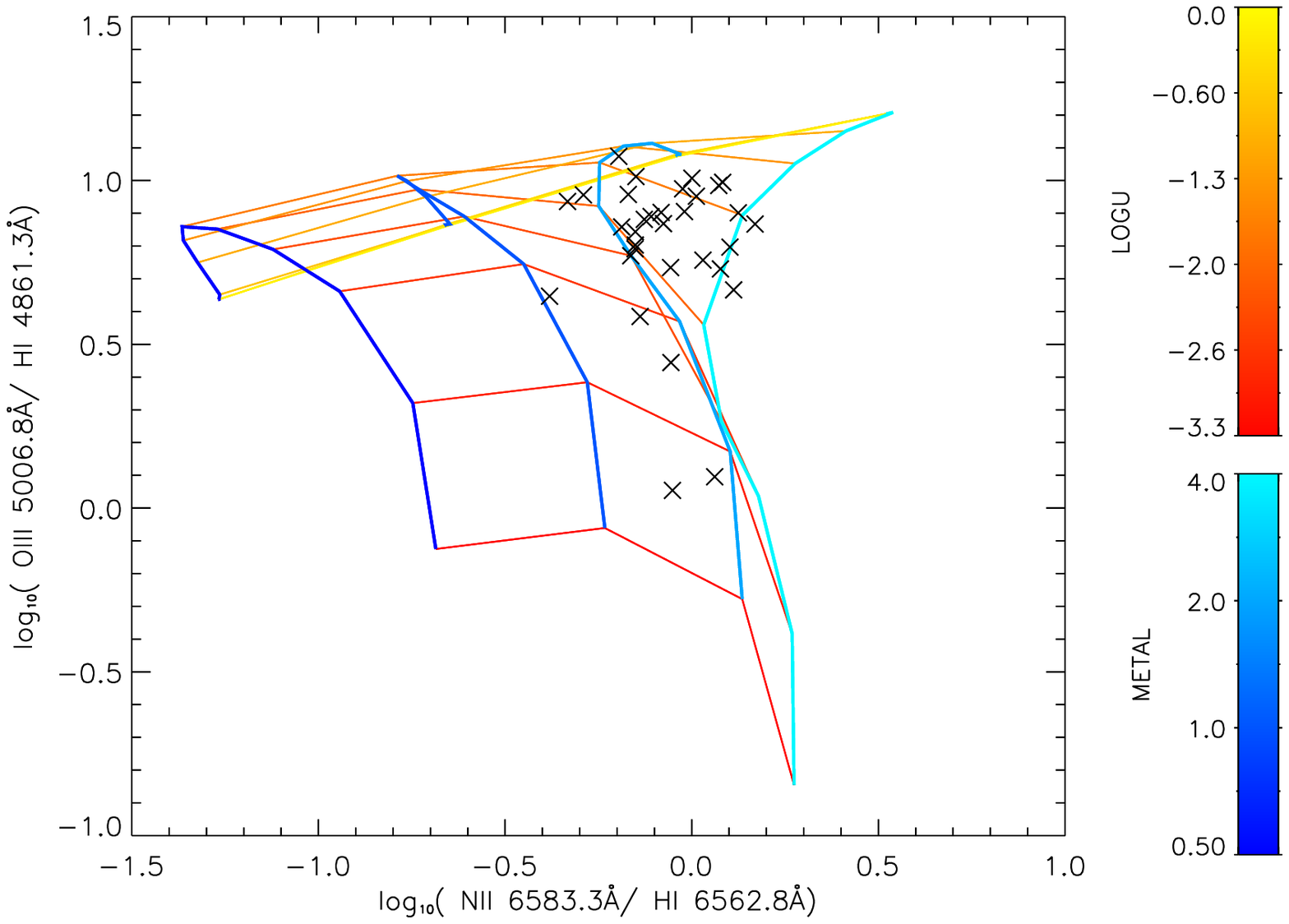}}
    \mbox{\hspace{0.5cm}}
    \subfigure[]{\includegraphics[scale=0.46]{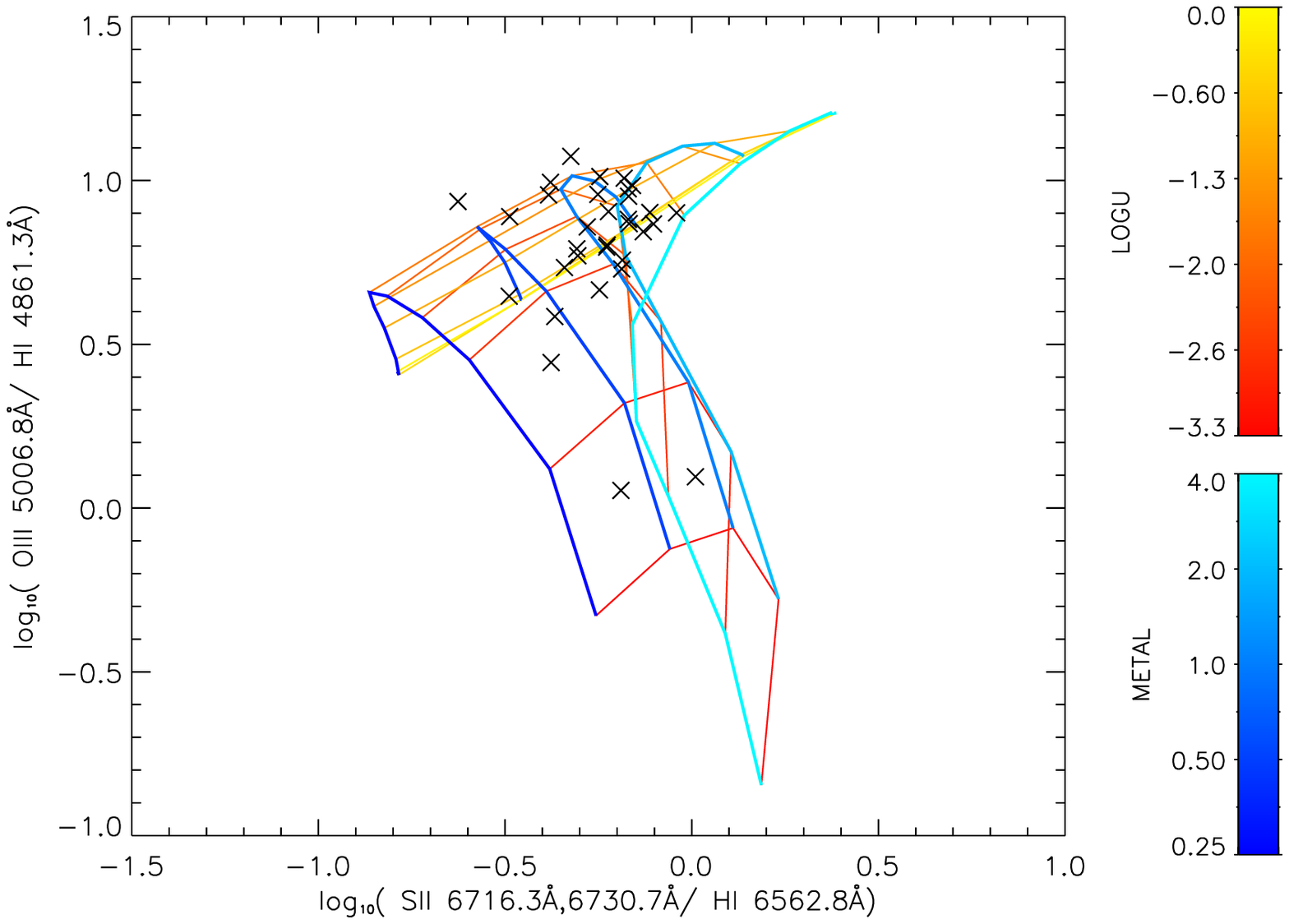}}
    \\
    \subfigure[]{\includegraphics[scale=0.46]{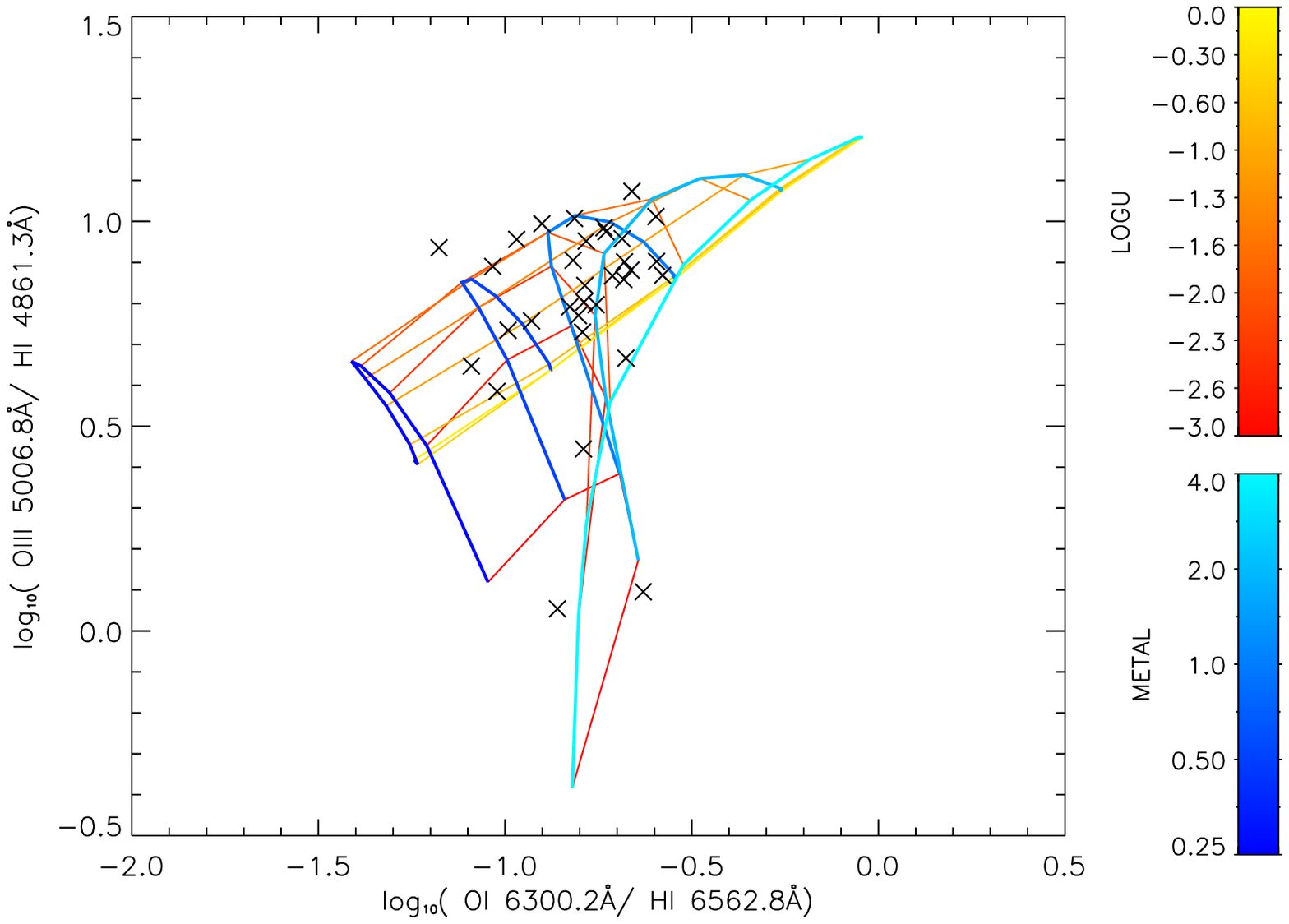}}
    \mbox{\hspace{0.5cm}}
    \subfigure[]{\includegraphics[scale=0.46]{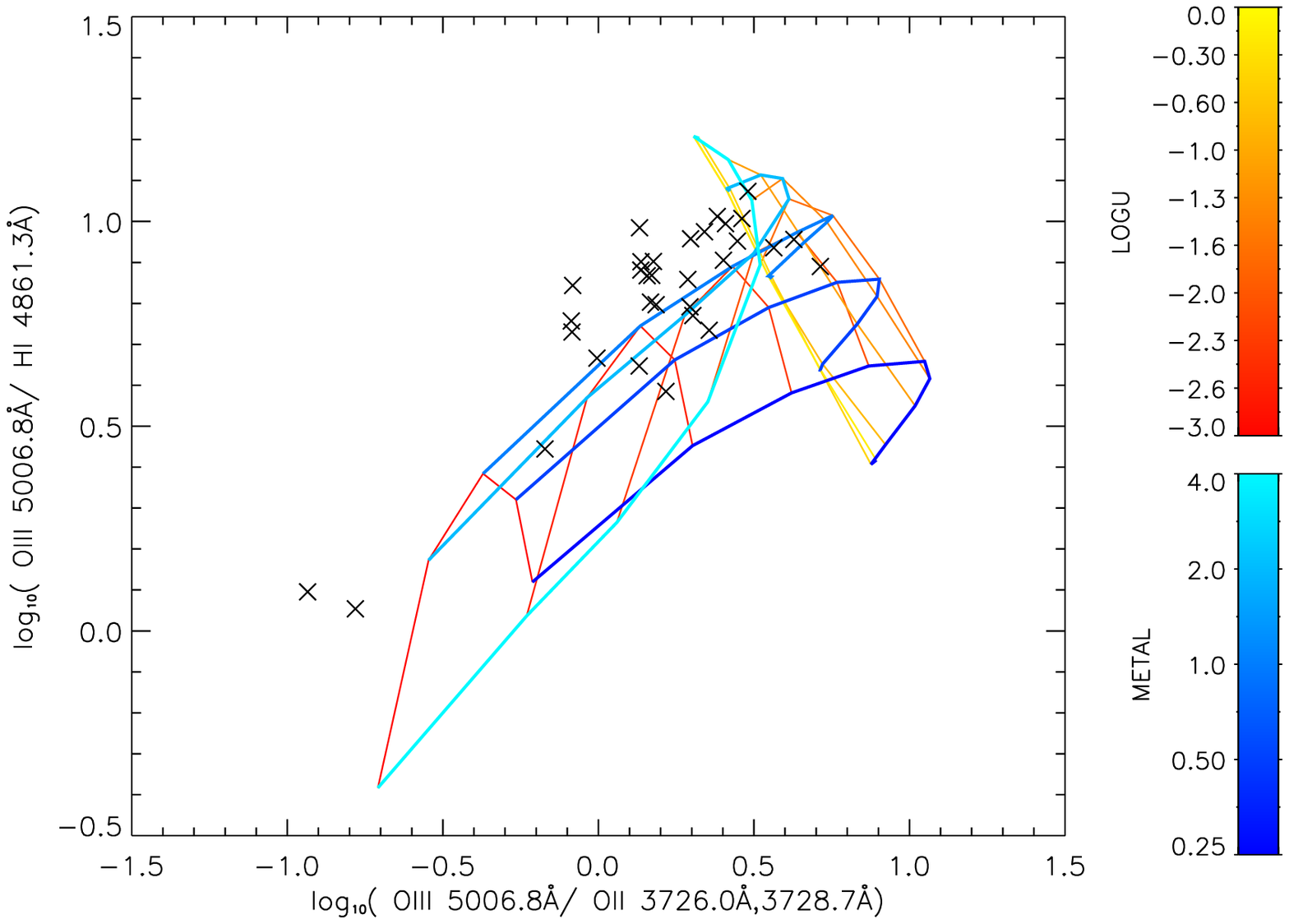}}
    \\
    \caption{Four panels show selected diagnostic diagrams in which models and observations are compared: (a) the [N II]/H$\alpha$ versus [O III]/H$\beta$ diagram; (b) the [S II]/H$\alpha$ versus [O III]/H$\beta$ diagram; (c) the [O I]/H$\alpha$ versus [O III]/H$\beta$ diagram ; (d) the [O III]/[O II] versus [O III]/H$\beta$ diagram. The observational data points (black crosses) are the NLR-dominated objects in A/BSY (i.e. sample A/B Seyferts) with $T_e$ $>$ 15 000 K. The model grids are plotted using \textsc{ITERA}, showing dusty, radiation-pressure dominated photoionized AGN models. We assume electron density of 100 cm$^{-3}$ and the power-law index $\alpha$ $=$ $-1.4$. The metallicity is in unit of solar metallicity.}
    \label{FigLrTem}
    \end{figure*}

    Fig. \ref{FigLrTem} shows the NLR observations on the line-ratio diagrams. Here we select four frequently used line-ratio diagnostic diagrams: [N {\small II}]/H$\alpha$ versus [O {\small III}]/H$\beta$, [S {\small II}]/H$\alpha$ versus [O {\small III}]/H$\beta$, [O {\small I}]/H$\alpha$ versus [O {\small III}]/H$\beta$ and [O {\small III}]/[O {\small II}] versus [O {\small III}]/H$\beta$. The observational data points are the NLR-dominated objects in A/BSY (i.e. sample A/B Seyferts) with $T_e$ $>$ 15 000 K (crosses). The model grids are plotted using \textsc{ITERA}\footnote{\textsc{ITERA}, and associated libraries and routines are all available from http://www.brentgroves.net/itera.html}, which is a \textsc{IDL} tool for emission-line ratio analysis (refer to \citealt{groves4} for details). The metallicities shown in the side bar are in units of solar metallicity.

    Temperature and ionization are fundamentally determined by the ionization source. First, we try both AGN (\citealt{groves5,groves6}) and shock \citep{allen} model libraries, which are currently available with \textsc{ITERA}. The former are photoionization models of the NLRs of AGN, while the latter consider fast shock effects. We find that the AGN models fit our observations better. Previous observations indicate that the NLR clouds are likely to be dusty and clumpy in nature. The dusty AGN models consider this effect to produce a physical model to explain why and how the AGN NLRs cluster on line ratio diagrams: dust dominates the opacity in dusty gas at high $U$, and the gas pressure gradient must match the radiation pressure gradient in isobaric systems. Therefore, these assumptions lead to a self-regulatory mechanism for the local ionization parameter and hence for the emission lines. We have conducted further experiments and found that the dusty, radiation-pressure dominated, photoionized AGN model grids can best reproduce our data when assuming a gas density of $n_H$$=$100 cm$^{-3}$ and a power-law index $\alpha$ $=$ $-1.4$. In Fig. \ref{FigLrTem}, we can see that the observed flux ratios are generally well described by the dusty AGN model predictions. However, some remaining problems still need to be addressed, as follows.

    \begin{enumerate}
      \item Our NLR data present a characteristic $n_e$ range of 10$^{2-3}$ cm$^{-3}$. The fact that the dusty AGN grids fit the data with density $n_H$ $=$ 100 cm$^{-3}$ better than other density constraints is illustrative.
      \item Despite the overall good fitting, the model grids present a systematic overestimation by a factor of 2 in the [O {\small III}]/[O {\small II}] versus [O {\small III}]/H$\beta$ diagram. [O {\small III}]/[O {\small II}] is sensitive to extinction correction and it decreased by approximately 30 per cent after correction. Even when including the uncertainties introduced by extinction corrections, the remaining discrepancy is still remarkable.
      \item Although emission line ratios support photoionization as the dominant ionization mechanism, the high gas temperatures and velocities observed within the NLR indicate that shocks might play a part in this region (see the discussion in Section 7.2 ).
      \item Some of these high-$T_e$, strong [O {\small III}]$\lambda$4363 emission Seyfert 2 galaxies show low-metallicity (i.e. Z/Z$_\odot$ $\sim$ 1; see the discussion in Section 7.3).
    \end{enumerate}

    To proceed, we will try to physically model the NLRs and probe these problems in a companion paper.

\section{M$_\star$ and SFR/M$_\star$}

    \begin{figure*}
    \centering
    \includegraphics[scale=0.95]{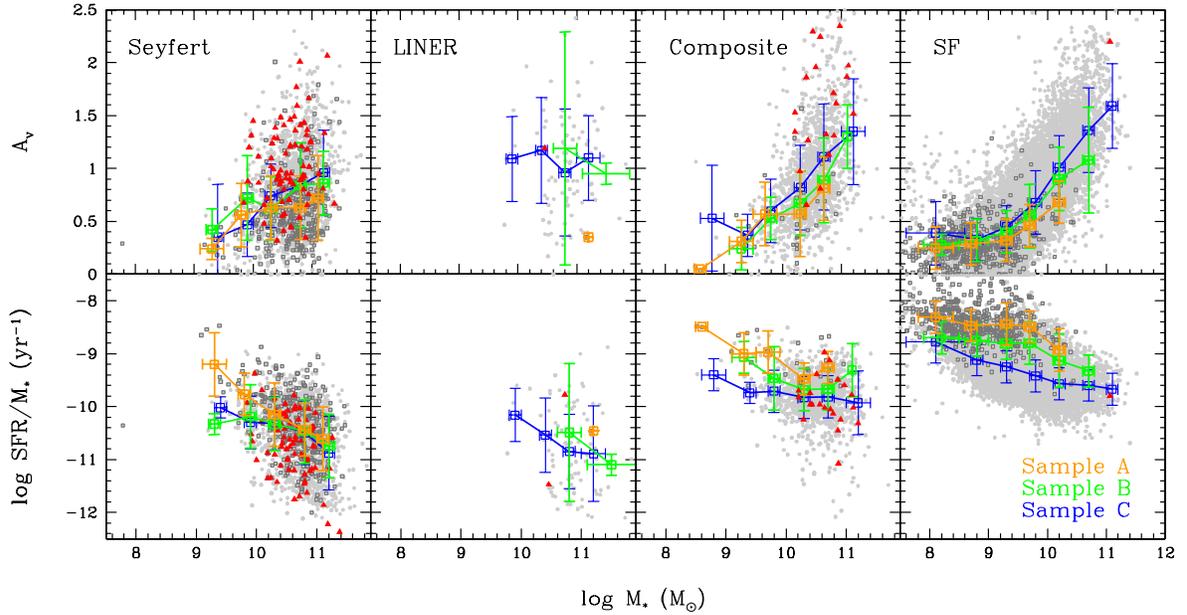}
    \\
    \caption{Stellar mass M$_\star$ versus dust attenuation $A_V$ (top) and specific star formation rate SFR/M$_\star$ (bottom) in bins of M$_\star$. Empty boxes are the median values of M$_\star$, $A_V$ and SFR/M$_\star$, with error bars showing 1$\sigma$ dispersions. From left to right: Seyferts, LINERs, composites and star-forming galaxies. Each type of galaxies is divided into sample A (orange), sample B (green) and sample C (blue). Red triangles are the NLR-dominated objects, dark grey open squares present the distribution of the sample A objects and grey background indicates the remaining objects (please refer to the online figure for details).}
    \label{FigMAvSFR}
    \end{figure*}

    The electron density and temperature are quantities representing the current statuses of NLRs, while stellar mass (M$_\star$) and star formation rate per unit mass (i.e. the specific star formation rate, SFR/M$_\star$) provide information about the galaxy formation histories. Either stellar mass is conserved or it is a slowly increasing quantity unaffected by morphological transformations and mergers (e.g. \citealt{hopkins1}). Traditionally, the masses of galaxies are estimated by dynamical methods from the kinematics of their stars and gas. The specific star formation rate has either explicitly or implicitly been used in numerous studies of field galaxies at z $<$ 1. It defines a useful measure of the rate at which new stars add to the assembled mass of a galaxy (i.e. a characteristic time-scale of star formation) and it is strongly related to several other important physical quantities (e.g. \citealt{clarke}).

    We adopt M$_\star$ and SFR/M$_\star$ derived using the methodologies of \citealt{kauffmann1} and \citealt{brinchmann}, respectively. \citet{kauffmann1} have constrained the stellar masses of galaxies using two stellar absorption-line indices (the 4000-{\AA} break strength D$_n$4000 and the Balmer absorption-line index H$\delta_A$). \citet{brinchmann} have obtained the SFR and SFR/M$_\star$ by two means: for star-forming galaxies, they used the observed H$\alpha$ luminosity and converted this into a SFR, while for composites and AGNs, they used a calibration based on D$_n$4000. The sample they used was drawn from SDSS DR2 with essentially all aperture bias removed. Note that their results are based on a model grid \citep{charlot}, ensuring a consistent picture for the attenuation of continuum and line emission photons.

    In this section, we turn our attention to M$_\star$ and SFR/M$_\star$ and we analyse both the relations between the two properties for different classes and their relations with the obtained plasma diagnostic results. Then, we compare these physical properties of the NLRs to those of the host galaxies.

\subsection{M$_\star$ versus SFR/M$_\star$}

    It is interesting to plot SFR/M$_\star$ as a function of M$_\star$. Previous studies have shown that galaxies generally split into two basic populations: concentrated galaxies with low SFR/M$_\star$ and low-mass, less concentrated galaxies with high SFR/M$_\star$ (e.g. \citealt{kauffmann2}; \citealt{brinchmann}). However, this is far from the full story. \citet{salim} have constructed a two-dimensional probability distribution function in the (log SFR/M$_\star$, log M$_\star$) plane and found that different types of emission-line galaxies occupy relatively distinct portions of the parameter space. Here, we attempt to provide more details.

    \begin{enumerate}
      \item Fig. \ref{FigMAvSFR} (bottom panels) shows that the SFR/M$_\star$ of the star-forming galaxies slowly decreases with increasing M$_\star$ from approximately 10$^8$ to 10$^{11}$ M$_\odot$, while those with high M$_\star$ show a much less pronounced correlation. Interestingly, those `less active' galaxies are indeed active galaxies: Seyferts and LINERs. Most composites are located at the junction connecting the AGN and star-forming branch. The distribution pattern agrees well with that presented by \citealt{salim}.
      \item As shown in Table \ref{Av_MassBin}, objects with high $A_V$ (e.g. sample C composites), generally reside in the conjoining region. They have a much higher $A_V$ than the Seyferts at constant M$_\star$. In general, high mass galaxies are older than low-mass galaxies, at a fixed stellar mass, studies show that high SFR galaxies yield relatively younger populations than those with lower SFR (e.g. \citealt{schaerer}). In other words, galaxies with lower SFR/M$_\star$ along with higher M$_\star$, on average, have experienced active star-forming activities and therefore are intrinsically older (e.g. \citealt{heavens}). Given that the amount of dust produced is proportional to the amount of stellar mass, as shown in the top panels in Fig. \ref{FigMAvSFR}, we speculate that some wide-reaching mechanisms (e.g. outflows driven by the AGN) might act to blow out dusty gases or consume a significant fraction of the dust content while lowering the SFR/M$_\star$ for massive galaxies.
      \item The selected NLR-dominated objects are plotted in Fig. \ref{FigMAvSFR} (red solid squares), contrasting with the background host galaxies (grey dots). In the (log M$_\star$, log SFR/M$_\star$) parameter planes, the NLRs distribute uniformly with the host galaxies, only biasing unsurprisingly to higher masses for the composites, because they reside in massive active galaxies by nature. To be observed, optical SFRs require emission-line luminosities and thus the aperture effects on emission-line luminosities can lead to strong biases in star-formation rate studies. In the (log M$_\star$, $A_V$) plane, we observe that the NLRs-dominated objects, on average, present higher $A_V$ values, which is also expected. \citet{brinchmann} have proposed an aperture correction method and have calculated the likelihood distribution of the SFR for a given set of spectra with global and fibre $g-r$ and $r-i$ colours. They have proposed that their correction method is robust, only if $\geqslant$ 20 per cent (a criterion argued by \citealt{kewley2}) of the total $r$-band light is sampled by the fibre; thus, a redshift z $>$ 0.04 is required for SDSS. However, as given in Section 5, only systems with ln$\phi$ $<$ 1 (i.e. z $<$ 0.046) are considered to be NLR-dominated. With these considerations, no precise conclusion can yet be reached, and further specific observations might be helpful.
      \item Finally, active galaxies with strong [O {\small III}] $\lambda$4363 emission behave unexpectedly. A/BSY and A/BL distribute uniformly with CSY/L, and A/BC are located on the high SFR/M$_\star$ side with lower M$_\star$ compared to CC objects (refer to Table \ref{Luminosity} for the abbreviations). Meanwhile, the situations are quite different in the star-forming cases: A/BSF definitely have both much lower M$_\star$ and higher SFR/M$_\star$ compared to CSF objects. As we have previously understood, strong [O {\small III}] $\lambda$4363 emissions in star-forming galaxies imply high $T_e$ and low cooling efficiency, and thus low gaseous metallicity; therefore, these objects have higher SFR/M$_\star$ than weak [O {\small III}] $\lambda$4363 emission star-forming galaxies at constant M$_\star$. However, this reasoning will not work in the AGN cases. Taking the Seyferts as an example, it seems that the occurrence of strong [O {\small III}] $\lambda$4363 emission is just a matter of frequency: there is no significant separation between the strong (i.e. S/N $>$ 5) and weaker (i.e. 1 $<$ S/N $<$ 5) [O {\small III}] $\lambda$4363 Seyferts in the (log M$_\star$, log SFR/M$_\star$) plane, compared to composites and star-forming galaxies. In addition, the strong [O {\small III}] $\lambda$4363 Seyferts distribute in a much wider range of SFR/M$_\star$ (i.e. -10.6 $<$ SFR/M$_\star$ $<$ -9.2) than the corresponding star-forming ones (i.e. -9 $<$ SFR/M$_\star$ $<$ -8.2). Assuming that the classifications of narrow emission-line galaxies are correct, we believe that such significantly different distribution patterns reveal some intrinsic physics related to [O {\small III}] $\lambda$4363 generating mechanisms and the properties of the central energy sources in galaxies. Some studies have shown that, in general, the ionization potential of the emission region is positively correlated with the [O {\small III}] $\lambda$4363 flux (e.g. \citealt{nagao}; \citealt{vaona}). Therefore, we refer to the results described above as the `evolutionary pattern of AGN with high ionization potential'. The key point is regarding the nature of the high-energy processes that excite the [O {\small III}] coronal lines in active galaxies. For example, the results from \citet{aird} demonstrate that the same physical processes regulate AGN activity in all galaxies in the M$_\star$ range (i.e. 9.5 $<$ log M$_\star$ $<$ 12) and could most likely be provoked by the energetically central nucleus.
    \end{enumerate}

    \begin{figure}
    \centering
    \includegraphics[scale=0.4]{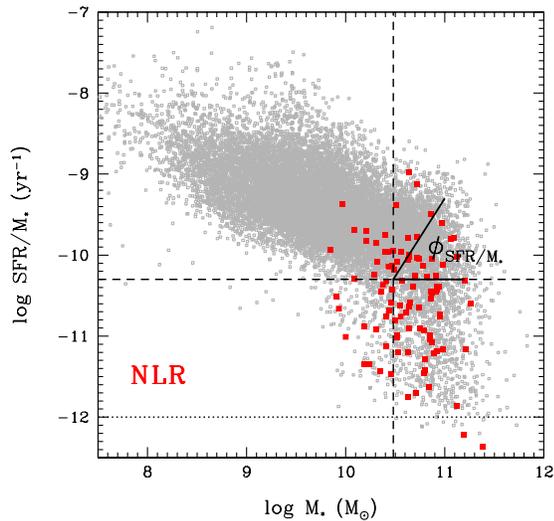}
    \\
    \caption{Sample distributions on the M$_\star$ versus SFR/M$_\star$ diagram, where red solid squares denote the NLR-dominated objects and grey squares indicate the remaining objects.  The dashed lines indicate log M$_\star$ = 10.48 (i.e. M$_\star$ = 3$\times$10$^{10}$ M$_\odot$ yr$^{-1}$, vertical) and log SFR/M$_\star$ = -10.30 (i.e. SFR/M$_\star$ = 5$\times$10$^{-11}$ yr$^{-1}$, horizontal). The dotted line presents log SFR/M$_\star$ = -12.}
    \label{FigMSFR}
    \end{figure}

    \begin{table}
     \caption{Summary of SFR/M$_\star$ in seven M$_\star$ bins.}
     \label{SFR_MassBin}
     \centering
        \begin{tabular}{l c c c c}
        \hline
         \\[-1.4 ex]
         log M$_\star$ (M$_\odot$) & \multicolumn{4}{c}{SFR/M$_\star$ (yr$^{-1}$)} \\
         & Seyferts & LINERs & Composites & Star-forming\\
         \\[-1.4 ex]
        \hline
         \\[-2 ex]
         8.1 & & & & -8.60 \\
         8.7 & & & -9.04 & -8.99 \\
         9.3 & -9.65 & & -9.36 & -9.17 \\
         9.8 & -10.16 & -10.16 & -9.66 & -9.57 \\
         10.3 & -10.28 & -10.54 & -9.81 & -9.56 \\
         10.7 & -10.48 & -10.83 & -9.80 & -9.60 \\
         11.2 & -10.80 & -10.89 & -9.91 & -9.70 \\
         \\[-2 ex]
        \hline
        \end{tabular}
    \end{table}

\subsection{Galaxy formation timescale sequence}

    \begin{figure*}
    \centering
    \includegraphics[scale=0.95]{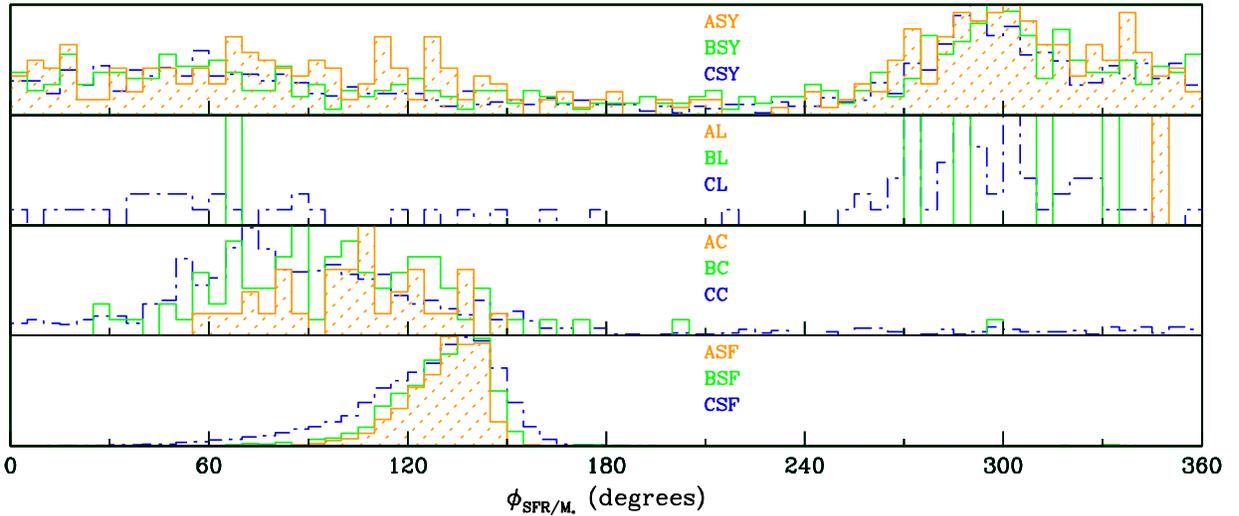}
    \\
    \caption{Normalized histograms of $\phi$$_{SFR/M_\star}$, which is the angle relative to the x-axis in the M$_\star$ versus SFR/M$_\star$ diagram, using the empirical point [-10.30, 10.48] as the vertex (see Fig. \ref{FigMSFR}). From top to bottom, we show Seyferts (A/B/CSY; refer to Table \ref{Luminosity} for the abbreviations), LINERs (A/B/CL), composites (A/B/CC) and star-forming galaxies (A/B/CSF). Blue dotted histograms show objects with 1 $<$ S/N $<$ 3 for [O {\small III}] $\lambda$4363, green dot-dashed ones indicate those with 3 $<$ S/N $<$ 5 and orange solid (shaded) histograms are those with S/N $>$ 5. Note that objects with log SFR $<$ -12 are not included.}
    \label{FigPhi}
    \end{figure*}

    It is now commonly accepted that the star formation history of a given galaxy depends strongly on its mass. Higher values of the specific SFR suggest that a larger fraction of stars was formed recently. The specific SFR is often called the galaxy present-day build-up time-scale, while M$_\star$ is the product of the overall galactic star-forming processes. We compare the mean SFR/M$_\star$ of the four classes in seven M$_\star$ bins. Table \ref{SFR_MassBin} clearly demonstrates that SFR/M$_L$ $\gtrsim$ SFR/M$_{SY}$ $<$ SFR/M$_C$ $<$ SFR/M$_{SF}$ at constant M$_\star$. Therefore, we consider these two clues combined as an indicator of galaxy formation history: it is generally true that $Y_{L}$ $\gtrsim$ $Y_{SY}$ $>$ $Y_{C}$ $>$ $Y_{SF}$, where $Y$ is present-day star-formation time-scale. However, this should not be confused with an actual galaxy age, because the specific SFR tells us only how long it would have taken to build a galaxy, assuming it had a current SFR throughout its lifetime.

    In Fig. \ref{FigMSFR}, we plot four classes of emission-line galaxies: Seyferts, LINERs, composites and star-forming in the (log M$_\star$, log SFR/M$_\star$) plane. We define an empirical base point on the logscale M$_\star$ versus SFR/M$_\star$ diagram, [-10.30, 10.48], which corresponds to M$_\star$ = 3 $\times$ 10$^{10}$ M$_\odot$ and SFR/M$_\star$ = 5 $\times$ 10$^{-11}$ yr$^{-1}$ (marked in Fig. \ref{FigMSFR}). The latter threshold is a `by-eye' division, because most star-forming galaxies perfectly distribute above the vertical dashed line. \citet{kauffmann4} have found that the transition from a young to old stellar population takes place at a characteristic M$_\star$ of 3 $\times$ 10$^{10}$ M$_\odot$, the logscale value of which is $\sim$10.48. For galaxies with M$_\star$, above this threshold, the distribution of sizes and concentrations at fixed stellar mass is independent of their local environments, and these galaxies have high stellar masses, high concentrations, high surface densities, little on-going star formation and red colours.

    Now we use the base point described above as the vertex to determine the angle, which is labelled $\phi_{SFR/M_\star}$, that each galaxy in our sample makes with the x-axis in the (log M$_\star$, log SFR/M$_\star$) plane. Fig. \ref{FigPhi} shows the normalized histograms of this angle corresponding to different galaxy types. Notice that, for accuracy, those with log-scaled specific SFR exceeding $-12$ are not included (i.e. objects that lie below the dotted lines in Fig.\ref{FigMSFR} are not counted because SFR/M$_\star$ estimates trend to be inaccurate past $\sim$10$^{-12}$ yr$^{-1}$; \citealt{brinchmann}). We can see that the histograms of the Seyferts and LINERs look quite similar and both peak at $\phi_{SFR/M_\star}$ $\sim$ 300$^\circ$, while composites and star-forming galaxies specify very different distributions: the former peaks at approximately 90$^\circ$ and the latter at approximately 140$^\circ$.

\subsection{Current status versus formation history}

\begin{figure*}
    \centering
    \includegraphics[scale=1]{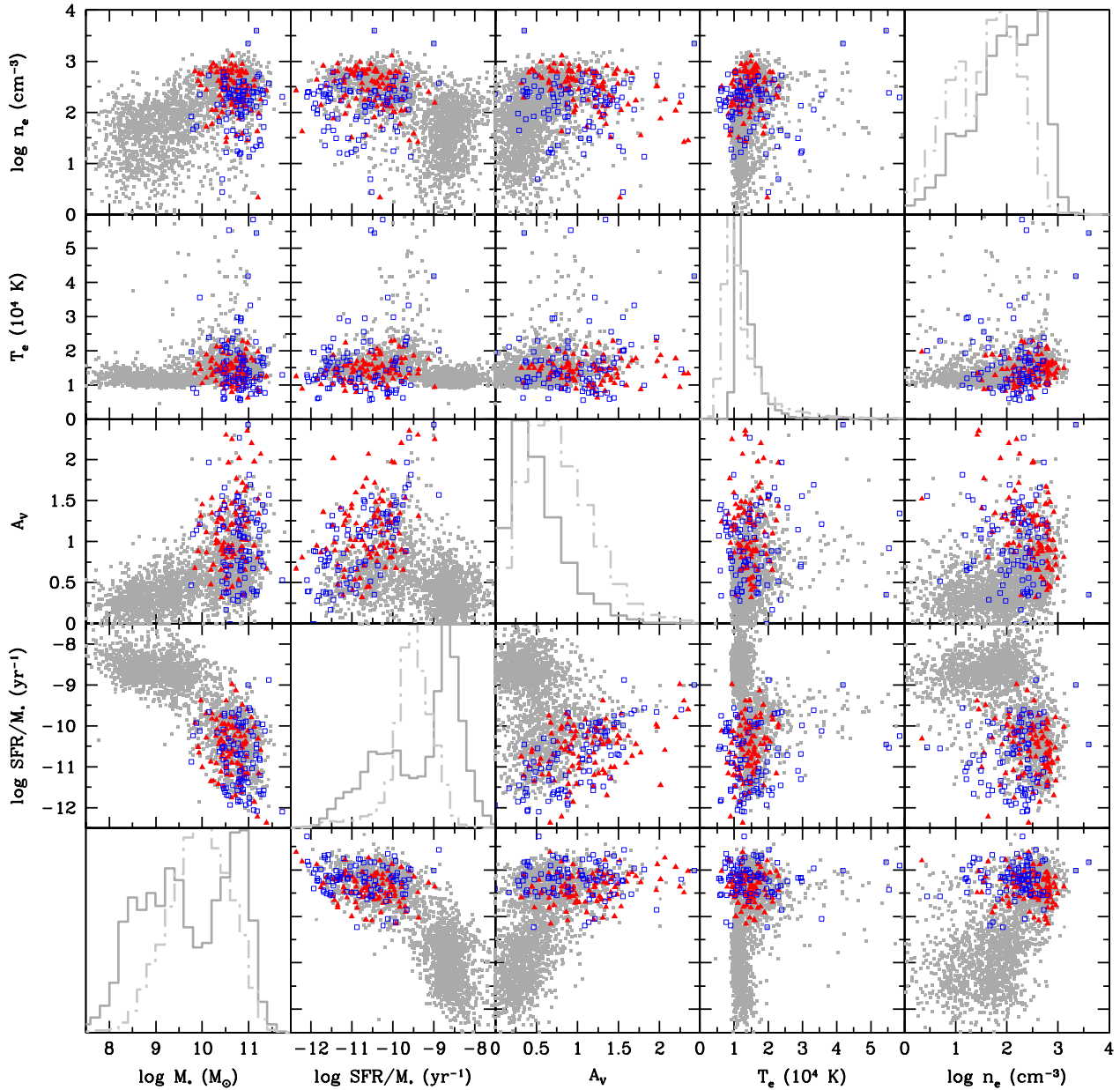}
    \\
    \caption{Distribution of galaxy properties. The diagonal panels show the normalized distribution of five properties independently: M$_\star$, SFR/M$_\star$, $A_V$, $T_e$ and $n_e$ (grey solid ones, sample A and sample B; lightgrey dot-dashed histograms, sample C. Objects from samples A and B present a bimodal distribution in M$_\star$, SFR/M$_\star$ is apparent, which is not seen in sample C. The off-diagonal panels show the bivariate distribution of each pair of properties, revealing the complex relationships among them. Red triangles are the NLR-dominated objects, blue open boxes mark out LINERs, while grey dots present objects from samples A and B.}
    \label{FigMSFRNeTe}
    \end{figure*}

    Utilizing previously obtained $n_e$ and $T_e$, which indicate the current statuses of galaxies, along with the dust attenuation $A_V$, finally we analyse their relations with galaxy formation history, indicated by M$_\star$ and SFR/M$_\star$. Fig. \ref{FigMSFRNeTe} illustrates our sample distributions in five property pairs. We notice the following.

    \begin{description}
      \item[\emph{Dust attenuation}.] The amount of dust content shows a pronounced correlation with the amount of stellar mass (i.e. massive galaxies contain more dust). As a function of SFR/M$_\star$, $A_V$ shows a positive correlation at low SFR/M$_\star$ and then becomes negative. Objects with log SFR/M$_\star$ $\sim$ -10 have the highest $A_V$, and the NLR-dominated objects (red triangles) have a wide range of $A_V$ and show both high M$_\star$ and low SFR/M$_\star$, compared to the host galaxies (grey dots).
      \item[\emph{Electron density}.] As a function of M$_\star$, $n_e$ also shows a positive correlation and reaches the highest value at the massive end. As a function of SFR/M$_\star$, a significant threshold is clearly revealed. A large fraction of objects with log SFR/M$_\star$ $<$ -10 have $n_e$ $>$ 100 cm$^{-3}$, while those with log SFR/M$_\star$ $>$ -10 show a much wider $n_e$ distribution. Most of the NLRs naturally lie in high $n_e$ regions.
      \item[\emph{Electron temperature}.] Compared to both $A_V$ and $n_e$, $T_e$ displays much less pronounced correlations with both M$_\star$ and SFR/M$_\star$. The existence of an unexpected high-$T_e$ branch is quite remarkable. The branch appears at approximately 9.5 $<$ log M$_\star$ $<$ 11 and -10.4 $<$ log SFR/M$_\star$ $<$ -9.4. The NLRs as a group show higher $T_e$ than the host galaxies.
    \end{description}

\begin{figure}
    \centering
    \includegraphics[scale=0.4]{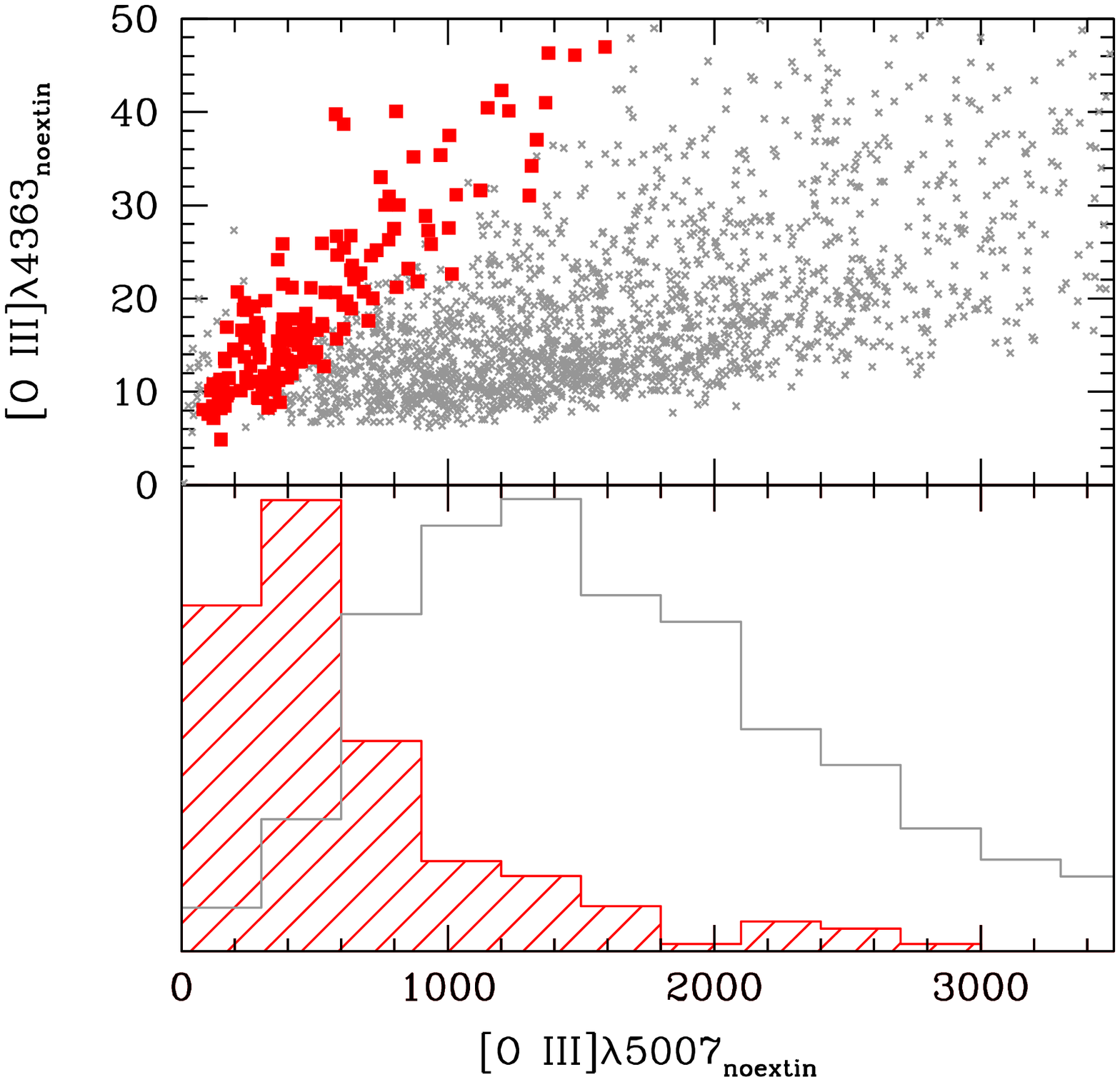}
    \\
    \caption{\emph{Top}: [O {\small III}] $\lambda$5007 versus [O {\small III}] $\lambda$4363 (pre-extinction correction fluxes in units of 10$^{-17}$ erg/s/cm$^2$). Red squares are objects with $T_e$ $>$ 2$\times$10$^4$ K and grey crosses present those with lower $T_e$. Bottom: normalized histograms of [O {\small III}] $\lambda$5007 fluxes for the above two groups. Note that only objects from samples A and B are plotted.}
    \label{FigO37}
    \end{figure}

    A significant $T_e$ `pumping-up' is shown in Fig. \ref{FigMSFRNeTe}. As listed in Table. \ref{HighTLowT}, while having similar redshifts, objects with $T_e$ $>$ 2$\times$10$^4$ K show much lower [O III] $\lambda$5007. Considering that $T_e$ increases rapidly with decreasing \textit{R}[O {\small III}] (i.e. \textit{I}[O {\small III}] $\lambda$5007/$\lambda$4363), we infer that the plume towards high $T_e$ is most likely the sign of an inaccurate $T_e$ estimate at the low \textit{R}[O {\small III}] end (Fig. \ref{FigO37}). Moreover, we can see in Fig. \ref{FigMSFRNeTe} that objects from sample A and sample B form a bimodal distribution in M$_\star$ and SFR/M$_\star$, which is not seen in sample C. Given that [O {\small III}] $\lambda$4363 correlates with galaxy activity, we argue that there exists a distinct low-activity population in active galaxies with 9.5 $<$ log M$_\star$ $<$ 11 and -10.4 $<$ log SFR/M$_\star$ $<$ -9.4, which might indicate some evolutionary stage.

    \begin{table}
    \centering
    \begin{threeparttable}
     \caption{Summary of high-$T_e$ objects ($>$ 2$\times$10$^4$ K) and low-$T_e$ objects (the rest) from samples A and B (median values). Note that [O III]$_{63}$ and [O III]$_{07}$ denote the pre-extinction correction fluxes of [O III] $\lambda$4363 and 5007, respectively (in units of 10$^{-17}$ erg/s/cm$^2$).}
     \label{HighTLowT}
        \begin{tabular}{l r r r r r}
        \hline
         \\[-1.4 ex]
         Objects & [O III]$_{63}$ & [O III]$_{07}$ & Redshift & M$_\star$ & SFR/M$_\star$ \\
         \\[-1.4 ex]
        \hline
         \\[-2 ex]
         High $T_e$ & 17 & 464 & 0.08 & 10.5 & -9.7 \\
         Low $T_e$ & 17 & 1683 & 0.07 & 9.5 & -8.9 \\
         \\[-2 ex]
        \hline
        \end{tabular}
    \end{threeparttable}
    \end{table}

\section{Discussion}

\subsection{Shock effects on $n_e$ and $T_e$}

    Stellar photoionization is not the only process that leads to the formation of emission lines. In fact, ionization can also be produced in shocks (e.g. \citealt{dopita2,dopita3}). To explain the observed emission line ratios, the possible contribution of shock-heating to photoionization must be examined. The cooling radiation from shocks produces high-energy photons that contribute to ionization. Kinetic energy is transformed into thermal energy, and the energetic thermal electrons can ionize the gas. Shocks are ubiquitous in galaxies. For example, shocks in AGNs can be caused by jets, winds or supersonic turbulence. Here, we find several emission line spectral indicators that can help trace the effects of shocks.

    \begin{enumerate}
     \item The density calculations show that the strong [O {\small III}] $\lambda$4363 AGN density is 100-150 cm$^{-3}$ higher than those that show no [O {\small III}] (see Section 4). Shocks generate compression, and therefore high gas densities. While temperature-sensitive line ratios change with the metallicity of the gas, they are also strongly affected by the ionization state of the gas. Therefore, shocks and the varying contribution of star formation and AGNs to the ionization become important.
    \item Table \ref{PhysicalConditions} (see column 12) lists objects with $T_e$ $>$ 20 000 K in each subsample, and we have shown in Section 5.3 that the dusty AGN model grids can fit some of these objects well. Shock heating produces very high temperatures (of the order of millions of K), leading to collisional ionization, and the production of highly ionized species. In fact, as we can see from Table \ref{PhysicalConditions} and Fig. \ref{FigMSFRNeTe}, high $T_e$ species are more easily found in LINERs and composites with strong [O {\small III}] $\lambda$4363 emission with certain SFR/M$_\star$ values. Thus, this indicates high-energy processes in their emission-line regions and the significant contribution of shocks evoked by the active nucleus, and even hot post-asymptotic giant branch stars and white dwarfs must be important (e.g. \citealt{stasinska}).
    \end{enumerate}

    \citet{groves0} have reported that the observed NLR spectra reveal much flatter mid-IR spectral slopes than those obtained in most NLR models. This suggests that shocks are a possible explanation for the difference between the model and observed slope. We confirm that it is both sensible and necessary to consider the combination of photoionization and shock. Additional multi-wavelength (especially infrared) observations could be very helpful in achieving this goal.

\subsection{Strong [O {\small III}] narrow-line AGN: low-metallicity candidates?}

    ``The physics of the NLR is simpler than that of the broad-line region; the physical conditions are in fact similar to H {\small II} regions, so that many of the metallicity-sensitive emission line ratios that are routinely applied to star-forming galaxies are also good diagnostics in the NLRs. \citet{storchi} have calibrated several of these emission line ratios for nearby AGN, using H {\small II} region determined metallicities and photoionisation models. However, the similarities between the NLRs and the H {\small II} regions in a galaxy lead to ambiguities when estimating abundances. Unless spatially resolved observations can be obtained, it is often difficult to disentangle the contributions to the emission-line spectrum from the NLR and star-forming regions within the galaxy", cited from \citet{groves2}.

    Low-metallicity AGNs appear to be very rare objects, at least locally. Currently, emission-line-based estimates of AGN metallicities at both high and low redshifts indicate that AGNs have predominantly solar-to-supersolar metallicities (e.g. \citealt{groves2}). This phenomenon is expected only if AGNs are old galaxies, are experiencing more supernova explosion events or even mergers, and are sitting on the later part of galaxy evolutionary sequence. From a theoretical point of view, when the total metallicity of a photoionized nebula decreases, there are several effects that cause changes in the final emission-line spectrum (e.g. \citealt{dopita1}; \citealt{osterbrock1}). For example, as metallicity decreases, the temperature of the nebula conversely increases. This is a result of the decrease in the efficient cooling metal emission lines and the availability of more high-energy photons to ionize the hydrogen (\citealt{sutherland}).

    Using the SDSS DR4 data, \citet{groves2} have found 40 low-metallicity AGN candidates selected out of the 23 000 Seyfert 2 galaxies. They expected that these candidates with masses below 10$^{10}$ M$_{\odot}$ could have metallicities approximately half those of typical AGN. Coincidentally, many of the candidates they found show the [O {\small III}] $\lambda$4363 line in their spectra. They have suggested that the use of a low-mass selection criterion supported by the observations reveals a correlation between mass- and metal-sensitive line ratios, and that the existence of [O {\small III}] $\lambda$4363 indicates higher gas temperatures in these objects and thus low metallicities. If their results are correct, we suggest that our strong [O {\small III}] $\lambda$4363 emission AGNs selected from SDSS DR7 similarly contain low-metallicity AGN candidates (as shown Fig. \ref{FigLrTem}). Using our refined ranges of density and temperature as input parameters for our NLR modelling, which will be the topic of a companion paper, we aim to reproduce the observational evidence and, as a definite challenge, obtain metallicity calibrations for active galaxies in order to further check these candidates.

\section{Summary}

    We have presented a statistical study based on spectroscopy of a large-sample from SDSS DR7. Our sample comprises 15 019 objects with a median redshift $\sim$ 0.08 and is divided based on [O {\small III}]$\lambda$4363 detection quality: 1100 in sample A (i.e. S/N $>$ 5), 1411 in sample B (i.e. 3 $<$ S/N $<$ 5), and 13 212 in sample C (i.e. 1 $<$ S/N $<$ 3). First, we determined electron density and electron temperature through \textit{I}[S {\small II}] $\lambda$6716 /$\lambda$6731 and \textit{I}[O {\small III}] $\lambda$5007/$\lambda$4363, respectively. To focus on the NLRs, we required that FWHM$_{Balmer}$ $>$ 300 km s$^{-1}$ and the physical aperture size ln$\phi$ $<$ 1 kpc to select the NLR-dominated objects within our sample. Then, we analysed a typical range of $n_e$ and $T_e$ of these objects and compared them with those of the host galaxies. Furthermore, we have utilized the plasma diagnostic results along with M$_\star$ and SFR/M$_\star$ to study the relations between these physical properties for four different classes of emission-line galaxies. Our main results are listed as follows.

    \begin{enumerate}
     \item The typical range of density in NLRs of AGNs is $10^{2-3}cm^{-3}$.
     \item The typical range of temperatures in NLRs of AGNs is 1.0-2.0$\times$10$^4$ K for Seyferts; for LINERs and composites, this range could be higher and wider.
     \item Transitions of both $n_e$ and $T_e$ from the NLRs to the discs are revealed (i.e. both $n_e$ and $T_e$ decrease outwards).
     \item Two sequences are proposed: $T_{LINER}$ $\gtrsim$ $T_{composite}$ $>$ $T_{Seyfert}$ $>$ $T_{star-forming}$ and $n_{LINER}$ $\gtrsim$ $n_{Seyfert}$ $>$ $n_{composite}$ $>$ $n_{star-forming}$.
     \item The median value of $n_e$ is $\sim150$ cm$^{-3}$ higher in both Seyferts and LINERs with strong [O {\small III}] $\lambda$4363 emission than in those with weak [O {\small III}] emission, most likely indicating some effect of shocks.
     \item LINERs and composites show the median $T_e$ to be approximately 2.0$\times$10$^4$ K, which is twice as much as that of star-forming galaxies and far too high to be explained by only stellar photoionization.
     \item Some strong [O {\small III}] $\lambda$4363 emission Seyfert 2 galaxies with $T_e$ $>$ 15 000 K can be fitted with dusty AGN model grids at low-metallicity (i.e. Z/Z$_\odot$ $\sim$ 1).
     \item LINERs with strong [O {\small III}] $\lambda$4363 emission simultaneously contain more dust and indicate the coexistence of old stellar populations and some specific high-energy process.
     \item Objects with high $A_V$ reside in a certain region of the M$_\star$ versus SFR/M$_\star$ diagram, suggesting that some wide-reaching mechanisms blow out or consume a significant fraction of dust contents, and meanwhile lower SFR/M$_\star$ for the massive galaxies.
     \item Considering characteristic present-day star-formation time-scale $Y$, we suggest the following relationship: $Y_{L}$ $\gtrsim$ $Y_{SY}$ $>$ $Y_{C}$ $>$ $Y_{SF}$.
     \item Seyferts with strong [O {\small III}] $\lambda$4363 emissions lie in the high M$_\star$ region and distribute uniformly with those showing weaker [O {\small III}] $\lambda$4363 emissions (the so-called evolutionary pattern of AGN with high ionization potential). The central nucleus might be responsible for the high ionization potential in Seyferts.
     \item The NLR-dominated objects have higher M$_\star$ and lower SFR/M$_\star$ than the host galaxies, most likely suggesting that the NLRs reside in old and massive active galaxies.
    \end{enumerate}

    In conclusion, we find several pieces of evidence pointing to additional source(s) of ionization other than the fact that photoionization by hot stars must be working to produce the highly ionized species such as the strong [O {\small III}] $\lambda$4363 line in NLRs of AGNs. Previous observations have shown correlations between jets and NLRs. The high gas temperatures and velocity dispersions present in our sample imply that the combination of shock and radiation heating might work as well. The best-known choice of ionization source is the energetic central nucleus, and (weak) shocks might be generated by feedback from AGNs in mergers or by the inflation of cavities by the central AGN. With our refined ranges of density and temperature as input parameters for the AGN NLR modelling, in a companion paper, we intend to reproduce the observational spectra and, furthermore, to obtain metallicity calibrations for active galaxies.

\section*{Acknowledgments}

    Zhitai Zhang dedicates her first research paper in astronomy to Professor Dame Jocelyn Bell Burnell who has played a leading role in our great understanding of the Universe. Her legend gives young minds a compass beyond the reach of science.

    Zhitai Zhang especially expresses her sincere thankfulness to Professor Gang Zhao (NAOC) for his gentle guidance.

    We gratefully acknowledge the anonymous referee for a careful and constructive revision that has improved this manuscript. The authors especially thank Jarle Brinchmann (Leiden University) for providing a number of detailed and crucial comments and suggestions, which have significantly enhanced this work. We also thank C. R. O'Dell (Vanderbilt University) and X. W. Liu (KIAA-PKU) for inspired pre-research discussions relating to the understanding of nebulae ionization structures, and L. Hao (SHAO) for sharing some interesting thoughts on the SDSS data.

    Funding for the creation and distribution of the SDSS-I and SDSS-II Archive has been provided by the Alfred P. Sloan Foundation, the Participating Institutions, the National Aeronautics and Space Administration, the National Science Foundation, the US Department of Energy, the Japanese Monbukagakusho and the Max Planck Society. The SDSS web site is http://www.sdss.org. Data products of SDSS DR7 have been publicly released by MPA/JHU group, a collaboration of researchers (current or former) from the MPA and the JHU, who have made their measured quantities on SDSS galaxies publicly available. This work was supported by the Natural Science Foundation of China (Grant Nos. 10933001 and 11273026).

    \bibliographystyle{mn2e}

\bsp

\label{lastpage}

\end{document}